\documentclass[a4paper, 11pt]{article}
\usepackage{comment} 
\usepackage[nointegrals]{wasysym}
\usepackage{fullpage} 
\usepackage{graphicx}
\usepackage{cite}
\usepackage{multicol}
\usepackage[perpage]{footmisc} 
\usepackage{sectsty}
\sectionfont{\large}
\usepackage{yhmath}
\usepackage{bm}
\usepackage{float}
\usepackage{amsmath}
\usepackage[utf8]{inputenc}
\usepackage{amssymb}
\usepackage{mathrsfs}
\usepackage{wrapfig}
\usepackage{enumitem}
\usepackage[english]{babel}
\usepackage{comment}
\usepackage{bbold}  
\usepackage[usenames,dvipsnames]{xcolor} 
\usepackage{xcolor,hyperref}
\usepackage{setspace}
\usepackage{abstract}
\usepackage{tablefootnote}

\usepackage[T1]{fontenc}
\usepackage[margin=0.94in]{geometry}
\def\Mpl{M_{\rm P}}
\def\Meff{M_{\rm eff}}
\usepackage{enumitem}

\begin{document}
\null\hfill IPMU25-0044 \\
\null\hfill{YITP-25-136}\\
\vspace*{\fill}

\vspace*{\fill}
\begin{center}
    \LARGE\textbf{{ \textcolor{Black}{The non-minimal 3-form cosmology and \\the rise of the cuscuton} }}

    \normalsize\textsc{Antonio De Felice$^{1}$, Anamaria Hell$^{2}$}
\end{center}

\begin{center}
$^{1}$ \textit{Center for Gravitational Physics and Quantum Information,\\
Yukawa Institute for Theoretical Physics,\\ Kyoto University,\\ 606-8502, Kyoto, Japan}\\
    $^{2}$ \textit{Kavli IPMU (WPI), UTIAS,\\ The University of Tokyo,\\ Kashiwa, Chiba 277-8583, Japan}
\end{center}
\thispagestyle{empty} 

\renewcommand{\abstractname}{\textsc{\textcolor{Black}{Abstract}}}

\begin{abstract}
We consider the 3-form theory with non-minimal coupling to gravity in an expanding Universe. First, we assume that the background is homogeneous and isotropic, and that the three-form is coupled to both the Ricci scalar and the Ricci tensor. We show that in this case, it propagates three degrees of freedom: a scalar mode and two tensor ones. Then, we consider an anisotropic background that corresponds to a Bianchi Type I Universe, and set the coupling with the Ricci tensor to zero. We show that, similarly to the Proca theory with non-minimal coupling to gravity, this case leads to two branches for the background solutions -- depending on the values of the 3-form. However, in contrast to the Proca case, we show that no extra modes appear.  We explore the no-ghost conditions and speed of propagation for all three modes in both branches. Finally, we show that one of the branches can be written as a theory of a constrained scalar, coupled to a cuscuton field. 
  \end{abstract}
 
\vfill
\small
\noindent\href{mailto:antonio.defelice@yukawa.kyoto-u.ac.jp}{\text{antonio.defelice@yukawa.kyoto-u.ac.jp}}\\
\href{mailto:anamaria.hell@ipmu.jp}{\text{anamaria.hell@ipmu.jp}}
\vspace*{\fill}

\clearpage
\pagenumbering{arabic} 
\newpage

\section{Introduction }

Models based on the 3-form fields are very curious among the massive gauge theories. If massive, they describe a single scalar \textit{degree of freedom (dof)}  in flat space-time. If massless, on the other hand, they have no propagating modes. Due to this interesting structure, they were considered in various physical frameworks. 

3-form fields naturally arise in string theory, super-symmetric models, and super-gravity \cite{Gates:1980ay, Gates:1980az, Binetruy:1996xw, Ovrut:1997ur, Bousso:2000xa, Feng:2000if, Nishino:2009zz, Bandos:2010yy, Bandos:2011fw, Groh:2012tf, Nishino:2013oea, Bielleman:2015ina, Morais:2016bev, Bouhmadi-Lopez:2016dzw, Buchbinder:2017vnb, Farakos:2016hly, Bandos:2016xyu, Aoki:2016rfz, Carta:2016ynn, Valenzuela:2016yny, Farakos:2017jme, Farakos:2017ocw, Kuzenko:2017vil,  Bandos:2018gjp, Becker:2017zwe, Herraez:2018vae, Cribiori:2018jjh, Nitta:2018vyc, Nitta:2018yzb, Bandos:2019qok, Lanza:2019xxg, Bandos:2019lps, Dudas:2019gxd, Cribiori:2020wch}. They were also key-ingredients in models that aim to account for the small values of the cosmological constant, and, in the presence of self-interactions, have been proposed to drive both the inflationary phase and the dark energy dominated expansion of the Universe  \cite{Hawking:1984hk, Brown:1987dd, Turok:1998he, Kaloper:2022jpv, Kaloper:2022oqv, Liu:2023vqp, Koivisto:2009fb, Koivisto:2009ew, Kobayashi:2009hj, Koivisto:2011rm,  DeFelice:2012jt,  Mulryne:2012ax,  Koivisto:2012xm, Kumar:2014oka, Barros:2015evi, Wongjun:2016tva, SravanKumar:2016biw, Morais:2017vlf,  BeltranAlmeida:2018nin}. Notably, due to their structure, which accounts for a different number of dof depending on the presence of mass, they can give rise to screening mechanisms and the strong coupling \cite{Barreiro:2016aln, Hell:2021wzm, Hell:2022wci, Hell:2025uoc}. More broadly, the 3-form fields were also considered in the context of black holes (including regular ones), wormholes, branes, stars, non-Gaussianities, and anisotropic manifolds \cite{ Urban:2012ib, SravanKumar:2016biw, Barros:2018lca, Barros:2020ghz, Barros:2021jbt, Bouhmadi-Lopez:2020wve, Bouhmadi-Lopez:2021zwt, Normann:2017aav, Bouhmadi-Lopez:2018lly, Almeida:2019xzt, Gordin:2023nsv}. They were also studied in relation to the naturalness problem \cite{Dvali:2005an, Dvali:2005zk, Giudice:2019iwl, Kaloper:2019xfj, Lee:2019efp, Lee:2019twi, Bordin:2019fek}.

Among the various 3-form field models, those with non-minimal couplings to gravity are particularly compelling. Like in broader p-form scenarios, such couplings can drive both early- and late-time cosmic acceleration \cite{Koivisto:2009sd, Germani:2009iq, Germani:2009gg}. However, they also significantly modify the structure of massive gauge theories. In Proca theory -- describing an Abelian massive, $U(1)$-gauge breaking, vector field
 -- non-minimal couplings can lead to non-trivial cosmological dynamics, but also introduce an additional degree of freedom beyond the scalar, two vector, and two tensor modes present in the minimally coupled case \cite{Golovnev:2009rm, Golovnev:2011yc, DeFelice:2025ykh}. While this extra mode is not necessarily a ghost, it was found to have vanishing propagation speed, indicating that the theory is strongly coupled \cite{DeFelice:2025ykh}. Moreover, problematic behavior can arise even when the field acts as a spectator with vanishing background value -- leading to instabilities, runaway modes, or strong coupling of tensor perturbations in flat spacetime, as shown for the Proca field\footnote{Note that the Generalized Proca Theory, in which the non-minimal coupling is given by the Einstein tensor avoids the instability, and does not give rise to extra dof \cite{Heisenberg:2016eld, DeFelice:2016yws}. } \cite{Himmetoglu:2009qi, Himmetoglu:2008zp,Golovnev:2009rm, Golovnev:2011yc, Capanelli:2024pzd, Hell:2024xbv}. 

The rich and sometimes unexpected behavior of massive vector fields with non-minimal coupling suggests that it is worthwhile to revisit other massive gauge theories with similar structure -- such as three-form fields. This motivation is further strengthened by the fact that a three-form is dual to a vector field, posing the natural question:  \textit{Is there a healthy theory of vector fields that admits general values of the couplings with the Ricci scalar and Ricci tensor? }

In this paper, we will revisit the cosmology of 3-form fields and investigate if such a model might give a positive answer to this question. In particular, by using the dual vector-field formulation, we will study the simplest cases of a 3-form field with non-minimal couplings to gravity, and demonstrate that while there are conditions under which such theories are well behaved, there are also branches where they might lead to problematic behaviors. 

We will first investigate the behavior of such fields in a homogeneous and isotropic expanding Universe, and show that in contrast to the Proca case, no additional modes arise for a general non-minimal coupling with both Ricci tensor and the Ricci scalar. Then, we will study the behavior of the 3-form theory in the anisotropic Bianchi Type I Universe, setting, for simplicity, the coupling to the Ricci tensor to zero. We will show that in this case, the theory shows behavior similar to the Proca case, in the sense that it will allow for two branches of the theory -- depending on the values of the vector field (or 3-form).  However, despite this similarity, we will show that the propagating number of modes does not change, and study the no-ghost conditions and speeds of propagation for the two branches.  

In general, studying the cosmological perturbation theory of massive gauge theories with non-minimal coupling in the Jordan frame can be non-trivial, with the highest obstacle being the number of terms that appear when one tries to write the theory containing only the propagating modes. To counter this, a better approach could be to try to write the theory in another way, where the behavior of the modes would be more manifest. 

In this light, we will show that one branch of 3-form theory with non-minimal coupling involving only the Ricci scalar can be written as a theory of a constrained scalar field with no kinetic term  that was recently {considered} in \cite{Hell:2025lgn},  coupled to the cuscuton field -- a scalar field with infinite speed of sound which propagates no dof on its own, thus depending on the dynamics other fields \cite{Afshordi:2006ad, Afshordi:2007yx}. {The constrained scalars naturally arise in the $f(R)$ theories of gravity, which can be brought by a conformal transformation to the form of the massive dilaton gravity -- a special case of the Brans Dicke theory with $\omega=0$ \cite{Brans:1961sx, Teyssandier:1983zz, OHanlon:1972xqa}. Notably, in contrast to having a non-minimal coupling with the Ricci tensor, and a general potential for the constrained scalar in these models, the theory in \cite{Hell:2025lgn} has potential restricted to a massive scalar field, giving a form similar to the variable gravity \cite{Wetterich:1987fm, Wetterich:1987fk, Wetterich:2013jsa} (but with a non-propagating scalar),  and also terms linear in the Ricci scalar and a cosmological constant. As a result, the dynamics does not depend on any value of the cosmological constant in the free theory. Moreover, it can drive the expansion of the Universe, admitting radiation-dominated, accelerated, and even super-Hubble phases without requiring external matter. Notably, this theory also admits solutions with phantom-like dark-energy, while being free of pathologies. 
}

 {The cuscuton field is also interesting on its own. }It propagates same number of degrees of freedom as General Relativity, but it is possible to distinguish it because of its features in the matter power spectra \cite{Afshordi:2007yx}. It belongs to the special case of the k-essence models \cite{Armendariz-Picon:1999hyi}, admits several cosmological solutions, with bouncing cosmologies, inflationary models, and anisotropic solutions \cite{Romano:2016jlz, Lin:2017fec, Boruah:2018pvq, Quintin:2019orx, Sakakihara:2020rdy, Kim:2020iwq, Dehghani:2025udv, Moghtaderi:2025cns}.  It was also considered in the scenarios with extra dimensions, where it was shown to admit accelerating solutions \cite{Ito:2019ztb}. Depending on its potential, it can also be considered as a low effective theory of the Ho\v{r}ava-Lifshitz theory \cite{ Afshordi:2009tt, Bhattacharyya:2016mah}, with the full Hamiltonian analysis performed in \cite{Gomes:2017tzd}. 
The intriguing properties of cuscuton have also opened an avenue to several new theories, including also the possibility to keep only the two tensor modes \cite{Cooney:2008wk, Chagoya:2016inc,  deRham:2016ged,  Boruah:2017tvg, Lin:2017oow,  Iyonaga:2018vnu, Pajer:2018egx, Gao:2019twq,  Grall:2019qof,Mukohyama:2019unx, DeFelice:2020eju, Iyonaga:2020bmm, Pookkillath:2021gdp, DeFelice:2022uxv, Ganz:2022iiv,Mylova:2023ddj, Bazeia:2025mzr}. {The combination of the cuscuton and the constrained scalar will additionally support the results on the number of degrees of freedom of the 3-form, with non-minimal coupling to the Ricci scalar, that we will find by evaluating the theory directly in the Jordan frame. }

The paper is organized as follows. In the section \ref{CP2}, we will study the 3-form with general non-minimal couplings to gravity, involving both Ricci scalar and the Ricci tensor, in the Friedmann-Lemaître-Robertson-Walker (FLRW) Universe. Then, in section \ref{CP3}, we will study the 3-form coupled only to the Ricci scalar in the anisotropic background. In section \ref{CP4}, we will show how one of the branches can be written to include a cuscuton field. In section \ref{CP5}, we will discuss the cosmological perturbations of this theory, and finally conclude in the section \ref{CP6}.

\section{The 3-form field in a homogeneous and isotropic universe}\label{CP2}

The theory of a 3-form, with  minimal coupling to gravity, is given by the following action: 
\begin{equation}\label{act3formmin}
    S=\int d^4x\sqrt{-g}\left[\frac{\Mpl^2}{2}\left(R-2\Lambda\right)-\frac{1}{48}W_{\mu\nu\alpha\beta}W^{\mu\nu\alpha\beta}+\frac{m^2}{12}C_{\mu\nu\alpha}C^{\mu\nu\alpha}\right],
\end{equation}
where,  
\begin{equation}
    W_{\mu\nu\alpha\beta}=C_{\nu\alpha\beta,\mu}-C_{\mu\alpha\beta,\nu}+C_{\beta\mu\nu,\alpha}-C_{\alpha\mu\nu,\beta},
\end{equation}
is the field-strength associated with the 3-form $C_{\mu\nu\rho}$. In the presence of mass, given by the last term of this action, this theory propagates a scalar mode, coming from the 3-form field, and two tensor modes that arise from the gravitational part of the action. On the other hand, if the mass term is absent, the 3-form contains no degrees of freedom (dof). 

The above theory can also be conveniently studied in its dual formulation, which can be obtained by substituting: 
\begin{equation}
    A_{\mu}=\varepsilon_{\mu\nu\rho\sigma}C^{\nu\rho\sigma},
\end{equation}
where $\varepsilon_{\mu\nu\rho\sigma}$ is the Levi-Civita tensor, and $A_{\mu}$ is a vector\footnote{One should note that if we treat the 3-form as a tensor, it would imply that $A_{\mu}$ is a pseudo-vector. However, the two formulations are in this context equivalent because all terms appear quadratic in the action. }. 
Then, the action (\ref{act3formmin}) becomes: 
\begin{equation}\label{actAmin}
    S=\int d^4x\sqrt{-g}\left[\frac{\Mpl^2}{2}\left(R-2\Lambda\right)+\frac{1}{2}\left(\nabla^{\mu}A_{\mu}\right)^2+\frac{m^2}{2}A_{\mu}A^{\mu}\right]. 
\end{equation}
As pointed out in \cite{Golovnev:2009rm, Golovnev:2011yc}, if one instead considers the presence of non-minimal couplings, new degrees of freedom might appear in massive gauge theories. 
This was also confirmed in \cite{DeFelice:2025ykh}, where Proca theory with non-minimal coupling was considered. In this paper, we will explore if the same holds also for the theory of a three-form with non-minimal coupling to gravity, which in its dual formulation is described by the following action: 
\begin{equation}\label{actnonMin}
    S=\int d^4x\sqrt{-g}\left[\frac{\Mpl^2}{2}\left(R-2\Lambda\right)+\frac{1}{2}\left(\nabla^{\mu}A_{\mu}\right)^2+\frac{m^2}{2}A_{\mu}A^{\mu}-\frac{1}{2}\left(\beta_1  RA_{\mu}A^{\mu}+\beta_2 R^{\mu\nu}A_{\mu}A_{\nu}\right)\right]. 
\end{equation}
In particular, in this section, we will explore the nature of this action in the homogeneous and isotropic universe. Then, in the following sections, we will give further general arguments that will support the results found in this section for the case of coupling with only the Ricci scalar. 

\subsection{The background configuration and perturbations}
Our first goal is to study the degrees of freedom corresponding to the action (\ref{actnonMin}) in the homogeneous and isotropic expanding Universe. For this, we will first find the equations for the background values of the metric and the vector field, and then perturb around these values. 

Let us assume the Friedmann–Lemaître–Robertson–Walker (FLRW) metric:
\begin{equation}
    ds^2=-N^2(t)dt^2+a^2(t)dx^idx^j,
\end{equation}
where $N(t)$ is the lapse, and consider the following background value for the vector field: 
\begin{equation}
    A_{\mu}=\left(N(t)A_0(t), 0, 0, 0 \right), 
\end{equation}
which preserves the isotropy of the space-time. 
By substituting the background values of the metric and the vector field into the action (\ref{actnonMin}), and varying with respect to the lapse, we find the constraint equation: 
\begin{equation}
    \begin{split}
      \left(\left(6 \beta_{1}+6 \beta_{2}-9\right) H^{2}-m^{2}\right) A_{0}^{2}+12 A_{0} \left(\beta_{1}+\frac{\beta_{2}}{2}-\frac{1}{2}\right)\dot{A}_0 H +6 H^{2} \Mpl^{2}-\dot{A}_0^{2}=\Lambda 2 a^{2} \Mpl^{2}
    \end{split}
\end{equation}

By varying the action with respect to the scale factor, and substituting the above constraint for the cosmological constant $\Lambda$, we further find: 
\begin{equation}
 \begin{split}
      0&=\left(\left(\frac{3}{2}-\beta_{1}-\beta_{2}\right) A_{0}^{2}-\Mpl^{2}\right) \frac{\ddot{a}}{a}-A_{0} \left(\beta_{1}+\frac{\beta_{2}}{2}-\frac{1}{2}\right)\ddot{A}_0\\
  &+H^{2} \left(-\frac{3}{2}+\beta_{1}+\beta_{2}\right) A_{0}^{2}+H \left(\beta_{1}-\frac{\beta_{2}}{2}+\frac{3}{2}\right) A_{0}+H^{2} \Mpl^{2}-\left(\beta_{1}+\frac{\beta_{2}}{2}\right)^{2}
 \end{split}
\end{equation}
Finally, by varying (\ref{actnonMin}) with respect to the temporal component of the vector field, we find: 
\begin{equation}
  A_{0} \left(\beta_{1}+\frac{\beta_{2}}{2}-\frac{1}{2}\right) \ddot{a}+a \left(-\frac{\ddot{A}_0}{6}+\left(\left(\beta_{1}+\frac{1}{2}\right) H^{2}-\frac{m^{2}}{6}\right) A_{0}-\frac{\dot{A}_0 H}{2}\right)=0
\end{equation}

The above three equations can be solved for the cosmological constant $\Lambda$, the second derivative of the scale factor $\Ddot{a}$, and the second derivative of the temporal component $\Ddot{A}_0$:
\begin{equation}\label{beom3form}
    \begin{split}
        \Lambda&=\frac{6 a^{2}}{2 a^{2}\Mpl^2}\left(\left(\left(-\frac{3}{2}+\beta_{1}+\beta_{2}\right) H^{2}-\frac{m^{2}}{6}\right) A_{0}^{2}+2 A_{0} \left(\beta_{1}+\frac{\beta_{2}}{2}-\frac{1}{2}\right)\dot{A}_0 H +H^{2} \Mpl^{2}-\frac{\dot{A}_0^{2}}{6}\right)\\
        \Ddot{a}&=\frac{a^{2}}{a \left(\left(6 \beta_{1}^{2}+\left(6 \beta_{2}-5\right) \beta_{1}+\frac{3 \beta_{2}^{2}}{2}-2 \beta_{2}\right) A_{0}^{2}+\Mpl^{2}\right)} \left(\left(-6 \beta_{1}^{2}+\left(-3 \beta_{2}+1\right) \beta_{1}-\frac{\beta_{2}}{2}\right) A_{0}^{2}H^{2}\right.\\&\left.+m^{2}A_{0}^{2} \left(\beta_{1}+\frac{\beta_{2}}{2}-\frac{1}{2}\right) +4 \left(\beta_{1}+\frac{\beta_{2}}{4}\right) H\dot{A}_0 A_{0}+H^{2} \Mpl^{2}-\left(\beta_{1}+\frac{\beta_{2}}{2}\right)\dot{A}_0^{2}\right)\\
        \Ddot{A}_0&=\frac{1}{a^{2} \left(\left(6 \beta_{1}^{2}+\left(6 \beta_{2}-5\right) \beta_{1}+\frac{3 \beta_{2}^{2}}{2}-2 \beta_{2}\right) A_{0}^{2}+\Mpl^{2}\right)
}\left(-3 \Mpl^{2} a^{2} H\dot{A}_0\right.\\&\left.-\left(-\frac{3}{2}+\beta_{1}+\beta_{2}\right) \left(\left(-12 \beta_{1}-3 \beta_{2}\right) H^{2}+m^{2}\right) a^{2} A_{0}^{3}+\frac{3 a^{2} H\dot{A}_0 \left(4 \beta_{1}^{2}-\beta_{2}^{2}+2 \beta_{1}+2 \beta_{2}\right) A_{0}^{2}}{2}\right.\\&\left.+\left(\left(-6 \left(\beta_{1}+\frac{\beta_{2}}{2}-\frac{1}{2}\right) \left(\beta_{1}+\frac{\beta_{2}}{2}\right)\dot{A}_0^{2}-m^{2} \Mpl^{2}\right) a^{2}+12 a^{2} H^{2} \Mpl^{2} \left(\beta_{1}+\frac{\beta_{2}}{4}\right)\right) A_{0}
\right)
    \end{split}
\end{equation}

The above equations take into account the evolution of the background values for the metric and the vector field. While in this work, we will not investigate the solutions to the above equations; when studying the dof, we will assume that the above equations are satisfied.

The perturbations of the metric and the vector field: 
\begin{equation}
    g_{\mu\nu}= g_{\mu\nu}^{(0)}+\delta  g_{\mu\nu},\qquad\text{and} \qquad A_{\mu}=A_{\mu}^{(0)}+\delta  A_{\mu},
\end{equation}
with, $g_{\mu\nu}^{(0)}$ and $A_{\mu}^{(0)}$ being the background values that satisfy (\ref{beom3form}), can be decomposed according to the irreducible representations of the rotation group into scalar, vector and tensor modes: 
\begin{equation}
    \begin{split}
        \delta g_{00}&=-2\phi\\
        \delta g_{0i}&=a(t)\left(S_i+B_{,i}\right)\\
        \delta g_{ij}&=a^2(t)\left(2\psi \delta_{ij}+2E_{,ij}+F_{i,j}+F_{j,i}+h_{ij}^T\right)\\
         \delta A_{0}&=A\\
         \qquad \delta A_i&=a(t)A_i^T+\chi_{,i}. 
         \label{eq:pert_metr_A}
    \end{split}
\end{equation}
Here, the comma denotes the derivative with respect to the spatial component. The vector and tensor modes satisfy: 
\begin{equation}
    S_{i,i}=0,\qquad F_{i,i}=0,\qquad  A_{i,i}^T=0,\qquad h_{ij,j}^T=0\qquad \text{and}\qquad h_{ii}^T=0. 
\end{equation}
At the linearized level, the scalar, vector, and tensor modes decouple, allowing us to study them separately.

\subsection{The scalar modes} 

Let us first consider the scalar modes. We will study the theory in the conformal gauge, 
\begin{equation}
    E=0\qquad\text{and}\qquad  B=0,
\end{equation}
in which the scalar potentials match with the gauge-invariant variables \cite{Mukhanov:2005sc, Bardeen:1980kt}. 
By expanding the action (\ref{actnonMin}) in terms of the scalar modes to the second order in perturbations, and performing several integrations by parts, we find the following Lagrangian density:
\begin{equation}\label{LagrdenFLRW}
    \begin{split}
\mathcal{L}&=c_1(t)\dot{\psi}^2+c_2(t)\dot{\phi}^2+c_3(t)\dot{A}^2+c_4(t)\dot{\phi}\dot{A}+c_5(t)\dot{\psi}\dot{A}+c_6(t)\dot{\psi}\dot{\phi}\\
        &+c_7(t)\dot{A}\psi+c_8(t)\dot{A}\phi+c_9(t)\dot{\phi}A+c_{10}(t)\dot{\phi}\psi+c_{11}(t)\dot{\psi}\phi+c_{12}(t)\dot{\psi}A+c_{13}(t)\dot{A}\Delta\chi\\
        &+c_{14}(t)\dot{\phi}\Delta\chi+c_{15}(t)\dot{\psi}\Delta\chi+c_{16}(t)\Delta\chi\Delta\chi+c_{17}(t)\chi\Delta A+c_{18}(t)\phi\Delta A+c_{19}(t)\psi\Delta A\\
        &+c_{20}(t)\chi\Delta \chi+c_{21}(t)\phi\Delta \chi+c_{22}(t)\phi\Delta \phi+c_{23}(t)\phi\Delta \psi+c_{24}\psi\Delta\psi\\
        &+c_{25}(t)A^2+c_{26}(t)\psi^2+c_{27}(t)\phi^2+c_{28}(t)\phi\psi+c_{29}(t)\psi A+c_{30}(t)\phi A
    \end{split}
\end{equation}
In the above expression, the time-dependent coefficients $c_i$, $i=1,...,30$ are given in Appendix A.1. 
We will express the fields in terms of the Fourier modes in the following way: 
\begin{equation}
    X(t,\Vec{x})=\int \frac{d^3k}{\left(2\pi\right)^{\frac{3}{2}}}X_k(t)e^{i\Vec{k}\Vec{x}}, \qquad\text{where}\qquad X=\left\{\phi,\psi,A,\chi\right\},
\end{equation}
and substitute the background equations of motion (\ref{beom3form}) into the Lagrangian density (\ref{LagrdenFLRW})\footnote{One should note that since we assume that the equations (\ref{beom3form}) are always satisfied, we will subsequently substitute them into the action at every step in the computation.}. 

We can notice among the above fields that the scalar $\chi$ is not propagating. In addition, by integrating by parts and  substituting  
\begin{equation}
    A_k=A_{2k}+\phi_kA_0,
\end{equation}
the scalar $\phi_k$ also stops propagating. To express the action only in terms of the physical degrees of freedom, we next find the constraint for the mode $\chi_k$, solve it, and substitute it back into the action. Followed by several integration by parts,  we then repeat the procedure with $\phi_k$ as well, writing the remaining action only in terms of the fields $\psi_k$ and $A_{2k}$. Then, we further substitute:
\begin{equation}
    A_{1k}=A_{2k}+\frac{\dot{A}_0}{H}\psi_k
\end{equation}
which renders the $\psi_k$ scalar to be constrained as well. By finding its constraint and substituting it back into the action, we find an action of only one scalar mode:
\begin{equation}
S=\int dtd^3k\left(b_1(t)\dot{A}_{2k}\dot{A}_{2(-k)}+b_2(t)A_{2k}A_{2(-k)}\right),
\end{equation}
where the coefficients $b_1$ and $b_2$ are in the limit of high momenta k given by: 
\begin{equation}\label{NGscalarsFLRW}
    \begin{split}
        b_1(t)&=\frac{3 a^{3} \dot{a}^{2} A_{0}^{2} \left(\beta_{1}+\frac{\beta_{2}}{2}\right) \left(\left(\beta_{1}^{2}+\left(-\frac{2}{3} \beta_{2}^{2}+\frac{5}{6} \beta_{2}\right) \beta_{1}-\frac{\beta_{2}^{3}}{3}+\frac{\beta_{2}^{2}}{2}\right) A_{0}^{2}+\Mpl^{2} \left(\beta_{1}-\frac{\beta_{2}}{6}\right)\right)}{\left(\beta_{1} A_{0}^{2}\dot{a}+a\dot{A}_0 \left(\beta_{1}+\frac{\beta_{2}}{2}\right) A_{0}+\Mpl^{2}\dot{a}\right)^{2}}\\
        b_2(t)&=-\frac{3 k^{2} A_{0}^{2} \dot{a}^{2} a \left(\beta_{1}+\frac{\beta_{2}}{2}\right) \left(\beta_{1}-\frac{\beta_{2}}{6}\right) \left(A_{0}^{2} \beta_{1}+\Mpl^{2}\right)}{\left(\beta_{1} A_{0}^{2} \dot{a}+a\dot{A}_0 \left(\beta_{1}+\frac{\beta_{2}}{2}\right) A_{0}+\Mpl^{2} \dot{a}\right)^{2}}. 
    \end{split}
\end{equation}
The positivity of $b_1(t)$ in the high-k limit gives us conditions that guarantee the absence of the pathological ghost dof:
\begin{equation}
    \text{NG1}=\beta_{1}+\frac{\beta_{2}}{2}\qquad\text{and}\qquad \text{NG2}=\left(\beta_{1}^{2}+\left(-\frac{2}{3} \beta_{2}^{2}+\frac{5}{6} \beta_{2}\right) \beta_{1}-\frac{\beta_{2}^{3}}{3}+\frac{\beta_{2}^{2}}{2}\right) A_{0}^{2}+\Mpl^{2} \left(\beta_{1}-\frac{\beta_{2}}{6}\right).
\end{equation}
In particular, if both $\text{NG1}$ and $\text{NG2}$ are positive, or both are negative, the scalar mode will be a healthy dof. In the high-k limit, its speed of propagation is given by:
\begin{equation}
    c_S^2=\frac{\left(A_{0}^{2} \beta_{1}+\Mpl^{2}\right) \left(6 \beta_{1}-\beta_{2}\right)}{\left(-2 \beta_{2}^{3}+\left(-4 \beta_{1}+3\right) \beta_{2}^{2}+5 \beta_{1} \beta_{2}+6 \beta_{1}^{2}\right) A_{0}^{2}+6 \left(\beta_{1}-\frac{\beta_{2}}{6}\right) \Mpl^{2}}
\end{equation}

Therefore, we can see that in the high-k limit, the scalar mode can be a healthy dof, provided that the above no-ghost conditions are satisfied, and the speed of propagation is larger than zero. It is important to notice that this mode is also present in the minimal theory, and thus, in contrast to the Proca field with non-minimal coupling to gravity, the non-minimal 3-form theory does not give rise to additional dof in the scalar sector. 

The above coefficients and the speed of propagation for the scalar mode were presented only in the high-k limit, due to the very complicated form of the actual expressions. Thus, to understand the nature of this mode further, let us point out some important cases, which can be derived by considering only the full expressions. 

\textsc{\textbf{Case 1}} In the absence of the Ricci tensor, $\beta_2=0$, the no-ghost condition, and the speed of propagation become: 
\begin{equation}
    A_0^2\beta_1+\Mpl^2>0\qquad\text{and}\qquad c_S^2=1. 
\end{equation}

\textsc{\textbf{Case 2}} If one sets in addition the coupling $\beta_1=0$, one finds an action with a kinetic term multiplied by the following function: 
\begin{equation}
    b_1(t)=\frac{m^{2} a^{5} \dot{a} \Mpl^{2}}{\left(\left(2 \Mpl^{2}-3 A_0^{2}\right) m^{2} a^{2}+2 k^{2} \Mpl^{2}\right) \dot{a}-2 \dot{A}_0 m^{2} A_0 a^{3}}.
\end{equation}
In the high-k limit, the above coefficient is always positive, and the mode propagates with a unit speed of propagation. However, the kinetic term is overall multiplied by mass, which suggests possible strong coupling when one considers the higher terms in the perturbation theory, in the limit when $m\to0$, as is the case when one considers instead the self-interacting 3-form theory in flat space-time \cite{Hell:2021wzm}.

\textsc{\textbf{Case 3}} Another important case is if the non-minimal coupling corresponds to the Einstein tensor. This is the case for 
\begin{equation}
    \beta_1=-\frac{\beta_2}{2},
\end{equation}
and in this case, we find again a non-vanishing action:
\begin{equation}
S=\int dtd^3k\left(\tilde{b}_1(t)\dot{A}_{2k}\dot{A}_{2(-k)}+\tilde{b}_2(t)A_{2k}A_{2(-k)}\right),
\end{equation}
with 
\begin{equation}
\begin{split}
        \tilde{b}_1&=\frac{\left(A_{0}^{2} \beta_{2}+2 \Mpl^{2}\right) a^{5} \left(-3 H^{2} A_{0}^{2} \beta_{2}^{2}+6 H^{2} \Mpl^{2} \beta_{2}-4 A_{0} \dot{A}_0 \beta_{2}^{2} H -m^{2} A_{0}^{2} \beta_{2}+2 \Mpl^{2} m^{2}\right)}{2 k^{2} \left(-A_{0}^{2} \beta_{2}+2 \Mpl^{2}\right)^{2}}\\
        \tilde{b}_2&=-\frac{a^{3} }{2 \left(A_{0}^{2} \beta_{2}+2 \Mpl^{2}\right) \left(-A_{0}^{2} \beta_{2}+2 \Mpl^{2}\right)^{2}}\left(16 A_{0}^{6} H^{2} \beta_{2}^{5}+32 \Mpl^{2} \beta_{2}^{4} A_{0}^{4} H^{2}-45 A_{0}^{6} H^{2} \beta_{2}^{4}\right.\\ & \left.-30 \Mpl^{2} \beta_{2}^{3} A_{0}^{4} H^{2}-24 A_{0}^{5} H \dot{A}_0 \beta_{2}^{4}-7 A_{0}^{6} \beta_{2}^{3} m^{2}+36 H^{2} \Mpl^{4} A_{0}^{2} \beta_{2}^{2}-32 \Mpl^{2} A_{0}^{3} H \dot{A}_0 \beta_{2}^{3}\right.\\ & \left.+6 \Mpl^{2} \beta_{2}^{2} A_{0}^{4} m^{2}+24 H^{2} \Mpl^{6} \beta_{2}-32 \Mpl^{4} A_{0} H \dot{A}_0 \beta_{2}^{2}+12 \Mpl^{4} m^{2} A_{0}^{2} \beta_{2}+8 \Mpl^{6} m^{2}\right)
\end{split}
\end{equation}

Therefore, the 3-form theory with non-minimal coupling to gravity in a homogeneous and isotropic expanding universe describes one scalar mode. Let us now consider the remaining dof -- the vector and tensor modes.

\subsection{The vector  modes} 
In the case of the vector mode perturbations, we will work in the Poisson gauge, in which 
\begin{equation}
    F_i=0. 
\end{equation}
To find the corresponding Lagrangian density, we expand the action (\ref{actnonMin}) up to the second order in perturbations. Then, after several integrations by parts, we find: 
\begin{equation}\label{actVec}
    \begin{split}
        \mathcal{L}=&d_1(t)\left(\dot{A}_i^TS_i+\dot{S}_iA_i^T\right)+d_2(t)S_i\Delta S_i+d_3(t)S_i\Delta A^T_i+d_4(t)S_iA_i^T+d_5(t)S_iS_i+d_6(t)A_i^TA_i^T.
    \end{split}
\end{equation}
In the above expression, the time-dependent coefficients $d_1 ... d_6$ are given in Appendix A.2. It is clear from the above Lagrangian density that both $S_i$ and $A_i^T$ do not propagate. In particular, by finding the constraint for one of them, solving it, and substituting it back into the action, the resulting Lagrangian density vanishes. Therefore, the non-minimal 3-form theory does not describe any vector modes. 
\subsection{The tensor  modes} 

Finally, let us consider the tensor modes. By expanding (\ref{actnonMin}) up to second order in tensor modes, and integrating by parts, we find: 
\begin{equation}\label{actTen}
    \begin{split}
    \mathcal{L}=&e_1(t)\dot{h}_{ij}^T\dot{h}_{ij}^T+e_2(t)h_{ij}^T\Delta h_{ij}^T+e_3(t)h_{ij}^T h_{ij}^T,
    \end{split}
\end{equation}
where the time-dependent coefficients $e_1, e_2$ and $e_3$ are given in Appendix A.3. By substituting also the background equations of motion, and writing the fields in the Fourier space: 
\begin{equation}
    h^T_{ij}=\sum_{\sigma=1,2}\int \frac{d^3k}{(2\pi)^{\frac{3}{2}}}\varepsilon_i^{T\sigma} h_k^{\sigma}(t)e^{i\Vec{k}\Vec{x}},
\end{equation}
with $\varepsilon_i^{T\sigma} $ being the polarizations, we find that the action associated with the tensor modes becomes: 
\begin{equation}
    \begin{split}
          S=\sum_{\sigma=1,2} \int dt d^3k \frac{a^{3}}{8}\left[ \left(\left(\beta_{1}+\beta_{2}\right) A_{0}^{2}+\Mpl^{2}\right)\dot{h}_k^{\sigma}\dot{h}_{-k}^{\sigma}-\frac{k^2}{8a^2} \left(A_{0}^{2} \beta_{1}+\Mpl^{2}\right) h_k^{\sigma}h_{-k}^{\sigma},
\right]
    \end{split}
\end{equation}
The first term in the above expression gives us the no-ghost condition:
\begin{equation}
    \left(\beta_{1}+\beta_{2}\right) A_{0}^{2}+\Mpl^{2}>0
\end{equation}
while the speed of propagation for the tensor modes is given by: 
\begin{equation}
    c_T^2=\frac{A_{0}^{2} \beta_{1}+\Mpl^{2}}{A_{0}^{2} \beta_{1}+A_{0}^{2} \beta_{2}+\Mpl^{2}}
\end{equation}
Curiously, we can notice that the tensor modes take exactly the same form as for the Proca field with non-minimal coupling \cite{DeFelice:2025ykh}. Moreover, when $\beta_2=0$, we find unit speed of propagation.  

\section{The anisotropic universe }\label{CP3}

In the previous section, we studied the 3-form theory with non-minimal coupling to gravity in the homogeneous and isotropic universe. In contrast to the Proca case \cite{DeFelice:2025ykh}, we have seen that the presence of non-minimal couplings does not introduce any extra degrees of freedom. Moreover, we have seen that the Ricci scalar introduces non-trivial no-ghost conditions, while keeping the speed of propagation to unity. The Ricci tensor, in contrast, distorts this speed for both scalar and tensor modes. 

One might still suspect that the above FLRW Universe is only a special case. In particular, there could be other backgrounds that could greatly change the situation. In this section, we will investigate this question for the anisotropic Bianchi Type I Universe, exploring if the number and behavior of the dof remain the same. 

To simplify the analysis, in the rest of the paper we will focus only on the non-minimal coupling with the Ricci scalar, and set $\beta_2=0$, considering the following action:
\begin{equation}\label{actnonMin2}
    S=\int d^4x\sqrt{-g}\left[\frac{\Mpl^2}{2}\left(R-2\Lambda\right)+\frac{1}{2}\left(\nabla^{\mu}A_{\mu}\right)^2+\frac{m^2}{2}A_{\mu}A^{\mu}-\frac{1}{2}\beta_1  RA_{\mu}A^{\mu}\right]. 
\end{equation}

Following the approach described in the previous section, let us now study the background values and the perturbations for the action (\ref{actnonMin2}). 

\subsection{The setup}
We will consider the Bianchi Type I Universe, which is given by the following background metric: 
\begin{equation}
    ds^2=-N^2(t)dt^2+a^2(t)dx^2+b^2(t)\delta_{ij}dx^idx^j, \qquad i=1,2, \qquad j=1,2
\end{equation}
Similarly to the analysis in \cite{DeFelice:2025ykh}, we will assume that the (pseudo) vector field has non-vanishing spatial and temporal components: 

\begin{equation}
    A_{\mu}=\left(-N(t)A_0(t), a(t)A_1(t), 0, 0\right). 
\end{equation}
Let us now first find the background equations of motion.  By varying the action (\ref{actnonMin2}) with respect to the lapse $N(t)$, and setting it subsequently to unity, we find the constraint equation:
\begin{equation}\label{constraintBT1}
    \begin{split}
        0&=\left(\left(-m^{2} A_{0}^{2}+m^{2} A_{1}^{2}-2 \Lambda \,\Mpl^{2}-\dot{A}_{0}^{2}\right) b^{2}-8 \left(-\dot{A}_{0} \left(\beta_{1}-\frac{1}{2}\right) A_{0}+A_{1} \dot{A}_{1} \beta_{1}\right) \dot{b} b\right.\\&\left. -2 \left(\left(-\beta_{1}+2\right) A_{0}^{2}+A_{1}^{2} \beta_{1}-\Mpl^{2}\right) \dot{b}^{2}\right) a^{2}-4 \left(\left(-\dot{A}_{0} \left(\beta_{1}-\frac{1}{2}\right) A_{0}+A_{1} \dot{A}_{1} \beta_{1}\right) b \right.\\&\left.+\left(\left(-\beta_{1}+1\right) A_{0}^{2}+A_{1}^{2} \beta_{1}-\Mpl^{2}\right) \dot{b} \right) \dot{a} b a -A_{0}^{2} \dot{a}^{2} b^{2}
    \end{split}
\end{equation}
In the following, we will always set $N=1$  in all of the expressions. 
By varying the action respect to the scale factor $a(t)$, we find: 
\begin{equation}\label{aeqBT1}
    \begin{split}
        0&=\left(\left(-m^{2} A_{0}^{2}+4 \left(\beta_{1}-\frac{1}{2}\right) \ddot{A}_0 A_{0}+\left(4 \beta_{1}-1\right) \dot{A}_{0}^{2}+m^{2} A_{1}^{2}-4 A_{1} \ddot{A}_1 \beta_{1}-4 \dot{A}_{1}^{2} \beta_{1}-2 \Lambda \,\Mpl^{2}\right) b^{2}\right.\\&\left.+\left(4 \ddot{b} \left(\beta_{1}-1\right) A_{0}^{2}+8 \dot{b} \dot{A}_{0} \left(\beta_{1}-1\right) A_{0}-8 \dot{A}_{1} A_{1} \dot{b} \beta_{1}-4 \ddot{b} \left(A_{1}^{2} \beta_{1}-\Mpl^{2}\right)\right) b \right.\\&\left.-2 \left(-\beta_{1} A_{0}^{2}+A_{1}^{2} \beta_{1}-\Mpl^{2}\right) \dot{b}^{2}\right) a^{2}+\left(\left(-4 \dot{a} \dot{A}_{0} A_{0}-2 \ddot{a} A_{0}^{2}\right) b^{2}-4 \dot{a} \dot{b} b A_{0}^{2}\right) a +A_{0}^{2} \dot{a}^{2} b^{2},
    \end{split}
\end{equation}
while by varying the action with respect to the scale factor $b(t)$, we find:  
\begin{equation}\label{beqBT1}
    \begin{split}
       0&= \left(\left(-m^{2} A_{0}^{2}+4 \left(\beta_{1}-\frac{1}{2}\right) \ddot{A}_0 A_{0}+\left(4 \beta_{1}-1\right) \dot{A}_{0}^{2}+m^{2} A_{1}^{2}-4 A_{1} \ddot{A}_1 \beta_{1}-4 \dot{A}_{1}^{2} \beta_{1}-2 \Lambda \,\Mpl^{2}\right) b\right.\\&\left. +2 \ddot{b} \left(\beta_{1}-2\right) A_{0}^{2}+4 \dot{b} \dot{A}_{0} \left(\beta_{1}-2\right) A_{0}-4 \dot{A}_{1} A_{1} \dot{b} \beta_{1}-2 \ddot{b} \left(A_{1}^{2} \beta_{1}-\Mpl^{2}\right)\right) a^{2}\\&+\left(\left(2 \ddot{a} \left(\beta_{1}-1\right) A_{0}^{2}+4 \dot{a} \dot{A}_{0} \left(\beta_{1}-1\right) A_{0}-4 \dot{A}_{1} A_{1} \beta_{1} \dot{a} -2 \ddot{a} \left(A_{1}^{2} \beta_{1}-\Mpl^{2}\right)\right) b \right.\\&\left. -2 \dot{a} \left(\left(-\beta_{1}+2\right) A_{0}^{2}+A_{1}^{2} \beta_{1}-\Mpl^{2}\right) \dot{b} \right) a +A_{0}^{2} b \,\dot{a}^{2}
    \end{split}
\end{equation}
Furthermore, we find the following equation for the temporal component: 
\begin{equation}\label{beomA0}
    \begin{split}
        0&=\left(\left(-A_{0} m^{2}-\ddot{A}_0\right) b^{2}+\left(4 \left(\beta_{1}-\frac{1}{2}\right) \ddot{b} A_{0}-2 \dot{A}_{0} \dot{b} \right) b +2 A_{0} \dot{b}^{2} \left(\beta_{1}+1\right)\right) a^{2}\\&-\left(\left(-2 \ddot{a} \left(\beta_{1}-\frac{1}{2}\right) A_{0}+\dot{a} \dot{A}_{0}\right) b -4 \beta_{1} A_{0} \dot{a} \dot{b} \right) b a +b^{2} \dot{a}^{2} A_{0},
    \end{split}
\end{equation}
while the spatial component satisfies the following equation: 
\begin{equation}\label{constraintA1}
    0=A_{1} \left(\left(m^{2} a -2 \ddot{a} \beta_{1}\right) b^{2}+\left(-4 \dot{b} \dot{a} \beta_{1}-4 \ddot{b} \beta_{1} a \right) b -2 a \,\dot{b}^{2} \beta_{1}\right).
\end{equation}
We can notice that the above constraint for the spatial component (\ref{constraintA1}) yields two branches for its solutions: 
\begin{equation}
    A_1(t)=0\qquad\text{and}\qquad A_1(t)\neq 0. 
\end{equation}
In the following, we will explore them separately, studying the perturbations around the background described by the above equations. In particular, we will study the perturbations around the metric tensor and the vector field: 
\begin{equation}
    g_{\mu\nu}= g_{\mu\nu}^{(0)}+\delta g_{\mu\nu}\qquad \qquad A_{\mu}=A_{\mu}^{(0)}+\delta A_{\mu},
\end{equation}
where $g_{\mu\nu}^{(0)}$ and $A_{\mu}^{(0)}$ are the background variables that satisfy equations (\ref{constraintBT1}) -- (\ref{constraintA1}). The perturbations for the metric and vector field can be decomposed, respectively, as: 
\begin{equation}\label{mBT1pert}
    \begin{split}
        &\delta g_{00}=2\phi\\
        &\delta g_{0x}=\omega_{,x}\\
        &\delta g_{oi}=v_i+B_{,i},\qquad v_i=\varepsilon_{ij}v_{,j}\\
        &\delta g_{xx}=a^2\psi\\
        &\delta g_{xi}=\lambda_{i,x}+\mu_{,xi}\qquad  \lambda_i=\varepsilon_{ij}\lambda_{,j}\\
        &\delta g_{ij}=b^2(2\tau \delta_{ij}+2E_{,ij}+h_{i,j}+h_{j,i}), \qquad h_i=\varepsilon_{ij}h_{,j}
    \end{split}
\end{equation}
and 
\begin{equation}\label{vfieldBT1pert}
    \begin{split}
        & \delta A_0=-A_0 A\\
        & \delta A_1=aA_1 \pi_{,x}\\
        & \delta A_i=-A_i^T+\chi_{,i}\qquad A_i^T=\varepsilon_{ij}v_{A,j}.
    \end{split}
\end{equation}
In the above relations, $\varepsilon_{ij}$ is the Levi-Civita symbol, with $\varepsilon_{yz}=1$, and the comma $,i$ denotes a derivative with respect to $\{y, z\}$. The modes $\phi, \omega, B, \psi, \mu, \tau, E, A, \pi, \chi$ are known as the even modes, and $v, \lambda, h, v_A$ are known as the odd modes. In linearized order, the two types decouple, which will allow us to study them separately.

\subsection{Case 1: The $A_1\neq0$ branch}

Let us first consider the branch with $A_1\neq0$. To find the set of background equations of motion that govern the evolution in this case, we first solve the constraint equation (\ref{constraintBT1}) for the cosmological constant $\Lambda$. Then, we substitute the resulting expression into equations for the scale factor $a$, (\ref{aeqBT1}), and the two equations for the vector components (\ref{beomA0}) and (\ref{constraintA1}). We then solve these three equations for $\ddot{a}, \ddot{A}_0$ and $\ddot{A}_1$. By substituting the resulting expressions into the equation for the scale factor $b(t)$, given in (\ref{beqBT1}), and solving it further for $\ddot{b}$, we find the system of equations of motion for the following variables:
\begin{equation}
    \{\; \Lambda, \;\ddot{a}(t),\; \ddot{b}(t), \; \ddot{A}_0(t),\; \ddot{A}_1(t)\; \}
\end{equation}
Similarly to the previous section, we will always substitute these expressions into the action when studying the perturbations. Let us now study the even and odd modes separately.  

\subsubsection{Even modes }
Let us first consider the even modes, for which it is convenient to work in the following gauge: 
\begin{equation}
    E=0,\qquad \tau=0,\qquad\text{and}\qquad \mu=0. 
\end{equation}
Similarly to the previous cases, to study these modes, we expand the action up to second order in perturbations. Then, we substitute the background equations of motion and integrate by parts. 
In the following, we will outline the steps in the procedure to find the action only in terms of the propagating degrees of freedom. For simplicity, we will work in the Fourier space, in which the modes can be written as: 
\begin{equation}\label{fourier}
    X=\int \frac{dkd^2q}{(2\pi)^{3/2}}X(t,k,q)e^{ikx+iq_iy^i}, \qquad y^i={y,z}
\end{equation}
where X stands for each of the modes. 
By studying the kinetic matrix, we can notice that substituting
\begin{equation}
    \omega=\omega_2+\frac{I\phi A_0a}{A_1k},
\end{equation}
and
\begin{equation}
    A=\tilde{A}-\frac{iA_1k\omega_2}{aA_0},
\end{equation}
makes the fields $\phi, \omega_2, B$ and $\chi$ non-propagating, after one performs several integrations by parts. This means that we can find their constraints, solve them, and substitute back into the action. We first apply this procedure to the $\chi$ mode, obtaining an action that describes six modes -- $B$, $\phi$, $\psi$, $\omega_2$, $\tilde{A},$ and $\pi$. Then, after several integrations by parts, we again notice that the determinant of the kinetic matrix is vanishing, with the scalar $B$ not propagating. We find its constraint, solve it, and substitute back into the action, which then becomes a function of five modes -- $\phi$, $\psi$, $\omega_2$, $\tilde{A},$ and $\pi$. By further substituting: 
\begin{equation}
    \tilde{A}=\bar{A}+\frac{\pi A_1^2 k I}{A_0^2},
\end{equation}
we find that the mode $\pi$ entirely drops from the action. Moreover, the $\omega_2$ mode loses its quadratic part, appearing in the action only linearly. By varying the action with respect to it, we find a constraint that is independent of $\omega_2$, and that can be solved for $\phi$, whose kinetic term is also absent. By resolving this constraint for $\phi$ and substituting it back into the action, we find that $\omega_2$ drops out from the whole expression, resulting in an action with just two propagating modes -- $\psi$ and $\bar{A}$. To study the conditions under which these modes are healthy, we will consider two limits, depending on the momenta along the anisotropic direction, and the homogeneous subspace of the background space-time. 

\begin{center}
    \textit{The no-ghost conditions}
\end{center}

In the high-k limit, the no-ghost conditions that are satisfied by the two propagating modes are given by:
\begin{equation}
    \begin{split}
        \text{NG1}&=\frac{3 \beta_{1}^{2} A_{0}^{4} q^{4} \dot{b}^{2}  a^{6} \left(-A_{0}^{2} \beta_{1}+\beta_{1} A_{1}^{2}-\Mpl^{2}\right)^{2}}{16 k^{4} {\left(\left(-A_{0}^{2} \beta_{1}+\beta_{1} A_{1}^{2}-\Mpl^{2}\right) \dot{b} +\beta_{1} b \left(-\dot{A}_{0} A_{0}+A_{1} \dot{A}_{1}\right)\right)}^{2}},
    \end{split}
\end{equation}

and
\begin{equation}
    \text{NG2}=\frac{q^{4} a^{5} \left(A_{0}^{2} \beta_{1}-\beta_{1} A_{1}^{2}+\Mpl^{2}\right) }{16 k^{4} b^{2}}
\end{equation}
We can notice that the first condition is always satisfied, i.e., one of the modes is always healthy because $\text{NG1}>0$. The second mode will be healthy as well if the following inequality is satisfied: 
\begin{equation}\label{ineqBT1em}
    A_{0}^{2} \beta_{1}-\beta_{1} A_{1}^{2}+\Mpl^{2}>0. 
\end{equation}

 In the high-q limit, we find the following no-ghost conditions: 
\begin{equation}
    \begin{split}
        \text{NG1}&=\frac{3 b^{4} \dot{b}^{2} a^{4} \beta_{1}^{2} A_{0}^{4} \left(A_{0}^{2} \beta_{1}-\beta_{1} A_{1}^{2}+\Mpl^{2}\right)^{2}}{16 {\left(\left(\beta_{1} \left(\dot{A}_{0} A_{0}-A_{1} \dot{A}_{1}\right) b +\frac{1}{2}\dot{b} \left(A_{0}^{2} \beta_{1}-\beta_{1} A_{1}^{2}+\Mpl^{2}\right)\right) a +\frac{1}{2}b \dot{a} \left(A_{0}^{2} \beta_{1}-\beta_{1} A_{1}^{2}+\Mpl^{2}\right)\right)}^{2}}
  \end{split}
\end{equation}

and 
\begin{equation}
    \begin{split}
         \text{NG2}&=\frac{3 b^{2} \dot{b}^{2} a^{3} \left(A_{0}^{2} \beta_{1}-\beta_{1} A_{1}^{2}+\Mpl^{2}\right)}{16 b^{2} \dot{a}^{2}+16 \dot{a} \dot{b} a b +16 \dot{b}^{2} a^{2}}
    \end{split}
\end{equation}
We can notice that, similarly to the high-k case, the first condition is always satisfied. The second, on the other hand, is satisfied only if the inequality (\ref{ineqBT1em}) holds.

\begin{center}
    \textit{The speed of propagation}
\end{center}
The expressions for the speed of propagation are very complicated. To study it, we will consider several discrete values for the parameters in the theory, ensuring that the no-ghost conditions are satisfied, and then calculate the speed of propagation. 
In particular, will consider the following three cases: 

\textbf{A) Small values of $\beta_1$, and away from the isotropy: }
\begin{equation}
    \left[a = 6, b = {\frac{1001}{
1000}}, \dot{a} = 3, 
\dot{b} = {\frac{3001}{1000}}, 
A_0 = 40, 
\dot{A}_0 = 34, 
A_1 = 5, 
\dot{A}_1 = -19, m = 1, 
\Mpl = 1, \beta_1 = 1\right]
\end{equation}

\textbf{B) Large values of $\beta_1$, and close to the isotropy: }
\begin{equation}
    \left[a = 1, b = {\frac{101}{100
}}, \dot{a} = 27, 
\dot{b} = 25, 
A_0 = 23, 
\dot{A}_0 = 3, 
A_1 = {\frac{1}{5}}, 
\dot{A}_1 = -19, m = 13, 
\Mpl = 1, \beta_1 = 56\right]
\end{equation}

\textbf{C) Very small values of $\beta_1$, and close to the isotropy: }
\begin{equation}
    \left[a = 10, b = {\frac{10001}{
1000}}, \dot{a} = 3, 
\dot{b} = {\frac{3001}{1000}}, 
A_0 = 40, 
\dot{A}_0 = 34, 
A_1 = 5, 
\dot{A}_1 = -19, m = 1, 
\Mpl = 1, \beta_1 = {\frac{1}{49}}\right]
\end{equation}

We find that all three cases show similar behavior.
\textcolor{Black}{Since we are considering the anisotropic Universe, we will distinguish between the speed of propagation along the anisotropic direction, and associated with the homogeneous and isotropic subspace. In particular, to find it, we solve the discriminant equation \cite{DeFelice:2025ykh}: 
\begin{equation}
\det[\omega^{2}A-2i\omega B^{T}-M]=0\,.
\end{equation}
for $\omega$, where $A$ is the kinetic matrix for our system, $B$ is the mixing matrix, and $M$ is the mass matrix. The group velocity, which we will refer to as the speed of propagation for modes along the $x-$direction, is then defined as 
\begin{equation}
    c_x=a\partial\omega/\partial k,
\end{equation} while for the isotropic homogeneous subspace it is defined as \begin{equation}
    c_{\rm yz}=b\partial\omega/\partial q. 
\end{equation}
To find the speeds of propagation, we solve the above discriminant equation for $\omega$, yielding four solutions $\omega_{1},\omega_2,\omega_3,\omega_4$. Then, we substitute these solutions into the definitions of the speeds of propagation and analyze different limits.
\\
In particular, in the $x-$direction,
{we find}:  
\begin{equation}
    c_{xe1,2}\sim\pm\frac{k}{q^2}\to\pm\infty\,,
\end{equation}
in the high-$k$ limit,\footnote{\textcolor{Black}{Since we are assuming a WKB approximation for the solutions, here by the high-$k$ limit, we actually mean $k/a\gg q/b\gg {\rm max}(\dot{a}/a,\dot{b}/b)$. Instead, by the high-$q$ limit, we mean $q/b\gg k/a\gg {\rm max}(\dot{a}/a,\dot{b}/b)$. The momenta are comparable when $k/a\simeq q/b\gg{\rm max}(\dot{a}/a,\dot{b}/b)$.}} while the same in the high-$q$ limit becomes: 
\begin{equation}
      c_{xe1,2}\sim\frac{k}{q} \to 0.
\end{equation}
{So for the high-$q$ modes, the propagation along $x$ leads to possible strong-coupling issues.} If the physical momenta are comparable, the speed of propagation for this mode becomes:
\begin{equation}
   c_{xe1,2}\sim\pm \frac{k}{\sqrt{\alpha k^2 + \beta q^2}},
\end{equation}
where $\alpha$ and $\beta$ are positive constants. Notably, if $q=\frac{b k}{a}$, we find:
\begin{equation}
   c_{xe1,2}=\pm\frac{\sqrt{2}}{2}.
\end{equation}
In contrast to this mode, the speed of propagation for the other one vanishes to the leading order: 
\begin{equation}
    c_{xe3,4}\sim 0\pm\mathcal{O}\left(\frac{q^6}{k^3}\right)
\end{equation}
in the high-k limit. 
Usually, one would instead have a constant in the leading order, such as, for example, in the case of the massive scalar field, as in this case, the dispersion relation becomes endowed with a mass. The appearance of the vanishing speed of propagation at the leading order implies that there is a possible strong coupling, meaning that the nonlinear terms might be of the same order as the linear terms. 
In the high-$q$ limit, the speed of propagation for this mode becomes: 
\begin{equation}
    c_{xe3,4}\sim\frac{k}{q}\to0\,,
\end{equation}
\\
{once again indicating a possible strong-coupling issue}. If the two physical momenta are comparable, one further finds the same result as for the other mode: 
\begin{equation}
   c_{xe3,4}\sim\pm \frac{k}{\sqrt{\alpha k^2 + \beta q^2}}. 
\end{equation}
In contrast to the speed of propagation associated with the $x-$direction, the speed of propagation in the homogeneous subspace is unity in the high-$q$ limit for all modes: 
\begin{equation}
    c^2_{yze1,...,4}=1,
\end{equation}
while for the high-$k$ limit, we find:
\begin{equation}
    c_{yze1,2}\sim\pm\frac{k^2}{q^3}
\end{equation}
for two modes, and 
\begin{equation}
    c_{yze3,4}\sim \pm q
\end{equation}
for the other two modes. {This means the speed of propagation in the isotropic/homogeneous subspace becomes strongly superluminal ($k^2/q^3\to\infty$ and $q\to\infty$) for the high-$k$ modes.}
Curiously, if one assumes that $q=\frac{b k}{a}$, then we find: 
\begin{equation}
    c_{yze1,2,3,4}=\pm\frac{\sqrt{2}}{2}.
\end{equation}
Therefore, overall, we find a vanishing speed of propagation for one of the modes for the group velocity along the $x$ direction, which indicates a possible strong-coupling case, meaning that the non-linear terms dominate. 
}

\begin{center}
    \textit{The $\beta_1\to0$ limit}
\end{center}

It is important to note that the NG1 expressions in both high-k and high-q limits vanish when $\beta_1$ is set to zero, {and even when one considers the full expressions.} 

This might yield a potential strong coupling behavior in the limit of small values of the coupling, once the non-linear terms are taken into account. 
To investigate this, it is best to consider the terms associated with the normalized kinetic matrix of the modes: 
\begin{equation}
    A_{ij}\Dot{F}_i\Dot{F}_j,\qquad\text{where}\qquad {F}_i=\{{\psi}, {\Bar{A}_n}\},
\end{equation}
where
\begin{equation}
    \Bar{A}=\frac{\bar{A}_n}{\beta_1}.
\end{equation}
Then, the kinetic matrix is finite when $\beta_1=0$. 
{By exploring the mass matrix for the normalized modes, we then find that it blows up to infinity when $\beta_1\to0$, which indicates that these modes can be integrated out. }

\subsubsection{Odd modes }

Let us next consider the odd modes. They can be found by expanding the action (\ref{actnonMin2}) up to second order in perturbations. In their case, it is convenient to work in the gauge: 
\begin{equation}
    h=0, 
\end{equation}
and write the remaining modes, $v, \lambda$, and $v_A$, in terms of the Fourier modes, following the previous convention given in the relation (\ref{fourier}). 

After several integrations by parts and imposing the background equations of motion, we can notice that $v$ and $v_A$ do not propagate. Moreover, $v_A$ completely drops out from the action, implying the existence of an accidental gauge symmetry, similarly to the behavior of the even mode $\pi$.  Then, we find the constraint for $v$, solve it, and substitute it back into the action. After integrating by parts, we obtain an action that is only a function of the odd-mode $\lambda$. This mode yields the following no-ghost condition: 
\begin{equation}
    \text{NG}=\frac{\left(A_0^{2} a^{2} \beta_1  +\Mpl^{2} a^{2}-A_1^{2} \beta_1  \right) q^{4} k^{2}}{4 a \left(b^{2} k^{2}+q^{2} a^{2}\right)}
\end{equation}
and the speed of propagation, which is unity in both the high-q and the high-k limit: 
\begin{equation}
    c_o^2=1.
\end{equation}
Moreover, if one examines the mass matrix for this mode, one can notice that it becomes divergent in the limit when $\beta_1\to0$. 

\subsection{Case 2: The $A_1=0$ branch }

In the previous subsection, we studied the branch of the theory in which the spatial part of the vector field does not vanish. Let us now consider the second branch, in which $A_1(t)=0$. In this case, we will then solve the background equations for 
\begin{equation}
    \{\; \Lambda, \;\ddot{a}(t),\; \ddot{b}(t), \; \ddot{A}_0(t) \; \}
\end{equation}
and always substitute them into the action, when studying the perturbations, which, similarly to the previous case, will be decomposed into even and odd modes according to  (\ref{mBT1pert}) and (\ref{vfieldBT1pert}), with now $\delta A_1$ component redefined as: 
\begin{equation}
    \delta A_1=a \pi_{,x}
\end{equation}
 so that it does not vanish.  Let us now consider the behavior of the even and odd modes, following a procedure similar to the previous case. 

 \subsubsection{The even modes} 
 In order to study the even modes, we expand the action up to the second order in perturbations. Then, we substitute the background equations of motion, write the modes in the Fourier space according to (\ref{fourier}), and perform several integrations by parts. By substituting 
 \begin{equation}
     A=\tilde{A}-\phi,
 \end{equation}
we then find that the modes   $\phi$, $\omega$, $B$, $\pi$, and $\chi$ are not propagating. We then proceed in a similar manner to the previous cases, by finding the constraint equations that the above modes satisfy by varying the action with respect to them, solving the constraints, and substituting back into the action. In particular, we first perform this procedure, the $\chi$ mode, followed by $B$, {$\pi$},  then $\omega$, and finally for $\phi$.  This leaves us with the action which is a function of only two even modes -- $\tilde{A}$ and $\psi$. 

After reaching the reduced Lagrangian for the perturbations, $\psi$, and $\tilde{A}$, we can find the kinetic matrix of the perturbations, whose elements $K_{ij}$ are defined as $\mathcal{L}=A_{ij}\dot q_i\,\dot q_j+\dots$ (the expression in the $\dots$ is not quadratic in the time derivative of the perturbation fields), where $\vec q=(\psi,\tilde{A})^T$. In the following, we determine two conditions, $g_1\equiv A_{22}$ and $g_2\equiv A_{11}A_{22}-A_{12}^2$ which determine the sign of the eigenvalues of the matrix $A_{ij}$. In particular, $A_{ij}$ is positive definite if $g_1>0$ and $g_2>0$.

In the high-$k$ regime, we get
\begin{align}
    g_1^{k\to\infty} &=\frac{3 \left(\beta_{1} A_{0}^{2}+ \Mpl^{2}\right) \beta_{1}^{2} \dot{b}^{2} b^{2} a A_{0}^{4}}{{\left(\left(\beta_{1} A_{0}^{2}+ \Mpl^{2}\right) \dot{b}+\beta_{1} b \dot{A}_0 A_{0}\right)}^{2}}\,,\\
    g_2^{k\to\infty} &=\frac{3 \beta_{1}^{2} \dot{b}^{2} A_{0}^{4} a^{6} q^{4} \left(\beta_{1} A_{0}^{2}+ \Mpl^{2}\right)^{2}}{16 k^{4} {\left(\left(\beta_{1} A_{0}^{2}+\Mpl^{2}\right) \dot{b}+\beta_{1} b \dot{A}_0 A_{0}\right)}^{2}}\,.
\end{align}
Instead, in the high-$q$ regime, we obtain
\begin{align}
    g_1^{q\to\infty} &=\frac{4 \left(\beta_{1} A_{0}^{2}+ \Mpl^{2}\right) \left(\dot{b}^{2} a^{2}+\dot{b} \dot{a} a b +b^{2} \dot{a}^{2}\right) \beta_{1}^{2} b^{2} a A_{0}^{4}}{{\left[\left(2 \beta_{1} b \dot{A}_0 A_{0}+\left(\beta_{1} A_{0}^{2}+\Mpl^{2}\right) \dot{b}\right) a +b \dot{a} \left(\beta_{1} A_{0}^{2}+\Mpl^{2}\right)\right]}^{2}}\,,\\
    g_2^{q\to\infty} &=\frac{3 \dot{b}^{2} \left(\beta_{1} A_{0}^{2}+\Mpl^{2}\right)^{2} a^{4} b^{4} A_{0}^{4} \beta_{1}^{2}}{4 \left[(\dot{b} a +\dot{a} b ) \beta_{1} A_{0}^{2}+2 A_{0} a b \dot{A}_0 \beta_{1}+(\dot{b} a +\dot{a} b ) \Mpl^{2}\right]^{2}}\,.
\end{align}
From the above, the non-trivial no-ghost condition reduces to $\Mpl^2+\beta_1 A_0^2>0$.

At this point, the reduced Lagrangian density (in the real field) for the two fields read as $\mathcal{L}=A_{ij}\dot{q}_i\dot{q}_j+B_{12}(\dot{q}_1q_2-\dot{q}_2q_1)-C_{ij}q_iq_j$, with $A_{12}=A_{21}$ and  $C_{12}=C_{21}$. By studying the propagation in the high-$k$ and high-$q$ regimes, we find analytically that both modes propagate luminally. In other words, the discriminant equation in the high-$k$ and high-$q$ regimes become proportional to $[\omega^2-(k/a)^2]^2$ and $[\omega^2-(q/b)^2]^2$ respectively.

 \subsubsection{The odd modes } 

Let us now consider the behavior of the odd modes when $A_1(t)=0$. For this, we expand the action up to second order in perturbations, set the gauge $h=0$, and substitute the background equations of motion. Similarly to the previous case, we find that $v_A$ is constrained, and satisfies a trivial equation of motion, which is of the following form 
\begin{equation}
    f(A_0(t), \beta_1, k, q) \;v_A=0\qquad\to\qquad v_A=0. 
\end{equation}
In addition, we find that $v$ does not propagate as well. We find its contrast, solve it, and substitute it back into the action. The resulting action thus becomes a function of only one more, $\lambda$, which has the following no-ghost condition, 
\begin{equation}
    \text{NG}=\frac{a\left(A_0^{2}  \beta_1  +\Mpl^{2} \right) q^{4} k^{2}}{4  \left(b^{2} k^{2}+q^{2} a^{2}\right)}
\end{equation}
and a unit speed of propagation, similarly to the first case.

\section{The interacting cuscuton }\label{CP4}

The non-minimal coupling to gravity can give rise to additional degrees of freedom if one considers theories of vector fields. However, this does not always have to be true. As we have studied in the previous section, the theory of a vector field that is dual to the three-form avoids the appearance of extra modes, in comparison with its minimal version. Nevertheless, one may wonder if the extra mode might reappear in different backgrounds. In this section, we will argue why this might not necessarily be the case by rewriting the theory in another way, which will give us a clearer view of the structure of the theory for one of the branches. 

The appearance of the additional degrees of freedom is usually associated with both the Ricci scalar and the Ricci tensor. The latter, in addition, modifies the dispersion relation, making the speed of propagation different from unity, as we have seen in section 2. However, for simplicity, let us set this term to zero. Then, the action is given by:  
\begin{equation}
    S=\int d^4x\sqrt{-g}\left[\frac{\Mpl^2}{2}(R-2\Lambda)+\frac{1}{2}\left(\nabla^{\mu}A_{\mu}\right)^2+\frac{\Meff^2}{2}A_{\mu}A^{\mu}\right],\label{eq:azAmu}
\end{equation}
where
\begin{equation}\label{effmass}
    \Meff^2=m^2-\beta_1 R. 
\end{equation}
Let us now introduce a scalar field: 
\begin{equation}
  S_1= \int d^4x\sqrt{-g}\left[\frac{\Mpl^2}{2}(R-2\Lambda)-\frac{1}{2}\varphi^2+\varphi\nabla^{\mu}A_{\mu}+\frac{\Meff^2}{2}A_{\mu}A^{\mu}\right] 
\end{equation}
By varying wrt $\varphi$, we find: 
\begin{equation}
    \varphi=\nabla^{\mu}A_{\mu}.
\end{equation}
Substituting this into $S_1$, we recover the initial action. By varying with respect to the vector field, provided that $M_{\rm eff}^2\neq0$,  we find:
\begin{equation}
    A_{\mu}=\frac{1}{\Meff^2}\nabla_{\mu}\varphi.
\end{equation}
In terms of the components $A_\mu$, excluding the case $M_{\rm eff}^2=0$, corresponds to excluding the solution $A_1\neq0$ in a homogeneous background, for which $\varphi=\varphi(t)$. Therefore, the theory described here does not have all the solutions of the original theory of Eq.\ \eqref{eq:azAmu}, but can be considered a subset of it. Substituting the above expression back into the action, we find: 
\begin{equation}\label{scfact}
    S_2=\int d^4x\sqrt{-g}\left[\frac{\Mpl^2}{2}(R-2\Lambda)-\frac{1}{2\Meff^2}\left(\nabla_{\mu}\varphi\nabla^{\mu}\varphi+\Meff^2\varphi^2\right)\right].
\end{equation}
We have then rewritten the previous action in terms of the scalar field. However, since the curvature is appearing inside the effective mass, it is not fully clear how to evaluate the degrees of freedom. For this, let us introduce another scalar field in the following way. First, we can write the action as: 
\begin{equation}
    S_3=\int d^4x\sqrt{-g}\left\{\frac{\Mpl^2}{2}(R-2\Lambda)-\frac{1}{2}\varphi^2-\frac{1}{2}\nabla_{\mu}\varphi\nabla^{\mu}\varphi\left[\frac{1}{m^2-\beta_1\rho}+\frac{\beta_1(R-\rho)}{(m^2-\beta_1\rho)^2}\right]\right\}.
\end{equation}
By varying with respect to $\rho$, we find: 
\begin{equation}
    \nabla_{\mu}\varphi\nabla^{\mu}\varphi\,\frac{\beta_1(R-\rho)}{(m^2-\beta_1\rho)^3}=0, 
\end{equation}
which implies that 
\begin{equation}
    \rho=R, \qquad\text{for}\qquad m^2-\beta_1\rho\neq 0\,,\quad{\rm and}\quad\nabla_{\mu}\varphi\nabla^{\mu}\varphi\neq0. 
\end{equation}
Since $\rho=R$ is algebraic for $\rho$, the above action reduces to the previous one when substituting this relation. At this point, let us perform a field redefinition for the field $\rho$. For this, let us further assume that the scalar field is time-like everywhere and anytime, therefore focusing only on a subset of the initial theory, and define: 
\begin{equation}
    \sigma^2=-\frac{\nabla_{\mu}\varphi\nabla^{\mu}\varphi}{(m^2-\beta_1\rho)^2}\,.
\end{equation}
This field redefinition should be thought of as an invertible relation between $\rho$ and $\sigma$, provided that $\nabla_\mu\varphi\nabla^\mu\varphi<0$.

 Then, as a consequence, we find that, if $\sigma\neq0$, that is, if $\nabla_\mu\varphi\nabla^\mu\varphi\neq0$, 
\begin{equation}
    \rho=\frac{m^2}{\beta_1}\mp \frac{\sqrt{X}}{\beta_1\sigma},
\end{equation}
where
\begin{equation}
    X \equiv -\nabla_{\mu}\varphi\nabla^{\mu}\varphi\,.
\end{equation}
At this point, we can insert $\rho$ into $S_3$ as an algebraic field redefinition. We find an equivalent two-branch Jordan-frame action, which can be rewritten as
\begin{align}
    \bar{S}_3^{\pm}&=\int d^4x\sqrt{-g}\left\{\frac{M_{\rm P}^2}{2}(R-2\Lambda)+\tfrac12\,\beta_1\sigma^2 R 
    \pm\sigma\sqrt{-\nabla_{\mu}\varphi\nabla^{\mu}\varphi}
    -\tfrac{1}{2}\varphi^2
    -\tfrac{1}{2}m^2\sigma^2
\right\}.\label{eq:action_int_cusc}
\end{align}
In the above, without loss of generality, we can include the $\pm$ inside the sign of $\sigma$, that is, we will rewrite $\pm\sigma\sqrt{X}\to\sigma\sqrt{X}$, so that the two branches corresponds to positive or negative values for $\sigma$, which cannot vanish, by assumption.
This action shows the presence of a cuscuton 
field coupled with $\sigma$, another scalar field that is non minimally coupled with gravity, {which lacks a kinetic term}. {The cosmology of the constrained scalar field was recently explored in \cite{Hell:2025uoc}, where it was shown that it gives rise to one scalar dof}. {Since the cuscuton usually does not give rise to new dof,} we expect to have only one additional scalar degree of freedom, the one coming from the scalar field $\sigma$, because of the non-minimal coupling with $R$. The presence of a timelike scalar field $\varphi$ anywhere and anytime leads to the presence of a preferred frame and does not allow solutions for which $\varphi$ is not timelike.

\subsection{The background equations of motion for a Bianchi I manifold}
By taking variations of the action (\ref{eq:action_int_cusc}) with respect to the lapse, and the background values of  $a$, $b$, $\varphi$, and $\sigma$, we find
\begin{align}
    \frac{m^{2} \sigma^{2}}{2}&=-\Lambda  \Mpl^{2}-\frac{\varphi^{2}}{2}+\frac{4 \dot{\sigma} \dot{b} \sigma  \beta_{1}}{b}+\frac{2 \dot{\sigma} \dot{a} \sigma  \beta_{1}}{a}+\frac{\dot{b}^{2} \sigma^{2} \beta_{1}}{b^{2}}+\frac{2 \dot{b} \dot{a} \sigma^{2} \beta_{1}}{a b}+\frac{\Mpl^{2} \dot{b}^{2}}{b^{2}}+\frac{2 \Mpl^{2} \dot{b} \dot{a}}{a b}\,,\\
   \frac{m^{2} \sigma^{2}}{2}&=2 \beta_1(\dot{\sigma}^{2} + \ddot{\sigma} \sigma)-\frac{\varphi^{2}}{2}+\sigma  \dot{\varphi}
   +\beta_1\sigma\left(\frac{4 \dot{\sigma} \dot{b}}{b}+\frac{2 \ddot{b} \sigma}{b}+\frac{\dot{b}^{2} \sigma}{b^{2}}\right)
   +\Mpl^2\left(\frac{2 \ddot{b}}{b}+\frac{\dot{b}^{2}}{b^{2}}-\Lambda \right),\\
   m^{2} \sigma^{2}&=-2 \Lambda  \Mpl^{2}-\varphi^{2}+4 \ddot{\sigma} \sigma  \beta_{1}+2 \sigma  \dot{\varphi}+4 \dot{\sigma}^{2} \beta_{1}+\frac{4 \dot{\sigma} \dot{b} \sigma  \beta_{1}}{b}+\frac{4 \dot{\sigma} \dot{a} \sigma  \beta_{1}}{a}+\frac{2 \ddot{b} \sigma^{2} \beta_{1}}{b}\\
   &+\frac{2 \ddot{a} \sigma^{2} \beta_{1}}{a}\nonumber+\frac{2 \dot{b} \dot{a} \sigma^{2} \beta_{1}}{a b}+\frac{2 \Mpl^{2} \ddot{b}}{b}+\frac{2 \Mpl^{2} \ddot{a}}{a}+\frac{2 \Mpl^{2} \dot{b} \dot{a}}{a b}\,,\\
    \varphi&= -\dot{\sigma}-\frac{2 \dot{b} \sigma}{b}-\frac{\dot{a} \sigma}{a}\,,\\
    \sigma  \,m^{2}&=\dot{\varphi}+\frac{4 \ddot{b} \sigma  \beta_{1}}{b}+\frac{2 \beta_{1} \sigma  \dot{b}^{2}}{b^{2}}+\frac{2 \ddot{a} \sigma  \beta_{1}}{a}+\frac{4 \dot{b} \dot{a} \sigma  \beta_{1}}{a b}\,.
\end{align}

We can formally solve the above equations of motion for $\Lambda$, $\ddot{a}$, $\ddot{b}$, $\ddot\sigma$, and $\dot\sigma$. For consistency, by taking the time derivative of $\dot\sigma$, after replacing the values of $\ddot\sigma$, we can find an equation for $\dot\varphi$, which reads
\begin{align}
    \dot{\varphi}&=\frac1{a^{2} \left(\left(6 \beta_{1}^{2}-5 \beta_{1}\right) \sigma^{2}+\Mpl^{2}\right) b^{2}}\,\bigl\{[(70 \dot{b}^{2} \beta_{1}^{2}-2 b^{2} m^{2} \beta_{1}) \sigma^{2}+40 \dot{b} \beta_{1}^{2} b \varphi  \sigma +b^{2} \left(m^{2} \Mpl^{2}+6 \beta_{1}^{2} \varphi^{2}\right)\nonumber\\
    &-4 \dot{b}^{2} \Mpl^{2} \beta_{1}]\, a^{2}
    -8 \left(-\tfrac{5}{2} \beta_{1} b \varphi  \sigma -\tfrac{15}{2} \dot{b} \beta_{1} \sigma^{2}+\dot{b} \Mpl^{2}\right) \dot{a} b \beta_{1} a +20 b^{2} \dot{a}^{2} \sigma^{2} \beta_{1}^{2} \sigma\bigr\}\,,
\end{align}
provided that $\left(6 \beta_{1}^{2}-5 \beta_{1}\right) \sigma^{2}+\Mpl^{2}\neq0$.

\subsection{Number of the propagating even degrees of freedom and no-ghost conditions for a Bianchi I manifold }
We will perturb the metric field as in Eq.\ \eqref{mBT1pert}. In addition, we introduce the following perturbation variables.
\begin{equation}
    \sigma=\sigma(t)+\delta\sigma\,,\qquad
    \varphi=\varphi(t)+\delta\varphi\,,
\end{equation}
and we will consider the following gauge fixing
\begin{equation}
    E=0\,,\qquad\mu=0\,,\qquad{\rm and}\qquad\delta\varphi=0\,,
\end{equation}
and we refer the reader to Eq.\ \eqref{mBT1pert} for the definition of the above fields.
This gauge choice, allowing for the unitary gauge (or the uniform field gauge), is consistent with the property that $\nabla_\mu\varphi\nabla^\mu\varphi<0$, so that $\dot\varphi\neq0$ on a homogeneous background.

At this point, we start performing a Fourier decomposition and several integrations by parts to integrate out all the auxiliary fields. After integrating out the field B by using its equation of motion, the field $\omega$ sets a constraint on the other fields as its quadratic term $\omega^2$ vanishes. We can use this equation as an equation for $\phi$ in terms of the remaining fields $\psi$, $\tau$, $\delta\sigma$, and their first time derivatives. The action is now a functional of the three fields $\psi$, $\tau$, and $\delta\sigma$. However, on making the following field redefinition
\begin{align}
    \tau&=\bar\tau +\tfrac12\,\frac{a}{\dot a}\,\frac{\dot b}b\,\psi\,,\\
    \delta\sigma&=\bar{\delta\sigma} -\tfrac12\,\frac{ab\varphi +2a\sigma\dot b + b \sigma \dot a}{b\dot a}\,\psi\,,
\end{align}
the field $\psi$ becomes a Lagrange multiplier that can be integrated out using its equation of motion.

Finally, the reduced action is only a functional of two propagating degrees of freedom, $\bar\tau$, and $\bar{\delta\sigma}$. We find two non-trivial no-ghost conditions in the high-$k$ regime:
\begin{align}
    &\frac{3 \left(\sigma^{2} \beta_{1}+\Mpl^{2}\right) \dot{b}^{2} a^{3} \sigma^{2} \beta_{1}^{2} b^{2}}{\left(-\beta_{1} \left(a \dot{b}+\dot{a} b \right) \sigma^{2}-\sigma  b \varphi  a \beta_{1}+\Mpl^2 \dot{b} a \right)^{2}}>0\,,\\
    &\frac{3 \beta_{1}^{2} \sigma^{2} a^{6} b^{2} \dot{a}^{2} q^{4} \left(\sigma^{2} \beta_{1}+\Mpl^{2}\right)^{2}}{4 \left(\beta_{1} \left(a \dot{b}+\dot{a} b \right) \sigma^{2}+\sigma  b \varphi  a \beta_{1}-\Mpl^{2} \dot{b} a \right)^{2} k^{4}}>0\,,
\end{align}
that leads to setting $\Mpl^2 + \beta_1\sigma^2>0$. In particular, this condition does not depend on the sign of $\sigma$. On expanding the no-ghost conditions in the high-$q$ regime, we find
\begin{align}
&\frac{4 \left(\sigma^{2} \beta_{1}+\Mpl^{2}\right) a \left(a^{2} \dot{b}^{2}+a b \dot{a} \dot{b}+b^{2} \dot{a}^{2}\right) \sigma^{2} \beta_{1}^{2} b^{2}}{{\left(\left(-2 \varphi  \beta_{1} \sigma  b +\left(-3 \sigma^{2} \beta_{1}+\Mpl^{2}\right) \dot{b}\right) a +b \dot{a} \left(-\sigma^{2} \beta_{1}+\Mpl^{2}\right)\right)}^{2}}>0\,,\\
&\frac{3 \dot{a}^{2} a^{2} \sigma^{2} b^{6} \left(\sigma^{2} \beta_{1}+\Mpl^{2}\right)^{2} \beta_{1}^{2}}{\left(\beta_{1} \left(-3 a \dot{b}-\dot{a} b \right) \sigma^{2}-2 \sigma  b \varphi  a \beta_{1}+\Mpl^{2} \left(a \dot{b}+\dot{a} b \right)\right)^{2}}>0\,,
\end{align}
two conditions that do not add any new constraint. 
\subsection{Diagonalization of the degrees of freedom}

From the previous analysis, we know that to avoid ghost degrees of freedom, we need to impose that $\Mpl^2+\beta_1\sigma^2>0$. We now consider this inequality to be fulfilled. In this case, to diagonalize the propagating degrees of freedom non-perturbatively, on any background (i.e.\ not only for Bianchi I manifolds), we can perform a conformal transformation in the form
\begin{equation}
    \bar{g}_{\mu\nu}=\Omega\,g_{\mu\nu}\,,
\end{equation}
for which we have (see the excellent paper by Dicke \cite{Dicke:1961gz})
\begin{equation}
    \sqrt{-g}=\Omega^{-2}\,\sqrt{-\bar g}\,,\quad
    g^{\mu\nu}=\Omega\,\bar{g}^{\mu\nu}\,,\quad
    R=\Omega\left[ \bar{R} +3 \bar\Box\ln\Omega-\frac32\,\bar{g}^{\mu\nu} \partial_\mu\ln\Omega \,\partial_\nu\ln\Omega  \right],
\end{equation}
where a bar stands for quantities/operators that are built only with $\bar{g}_{\mu\nu}$. Since $\Mpl^2+\beta_1\sigma^2>0$, we have defined
\begin{equation}
    \Omega\equiv\frac{\Mpl^2 +\beta_1\sigma^2}{\Mpl^2}>0\,.
\end{equation}
In this case, Eq.\ \eqref{eq:action_int_cusc} reduces to
\begin{align}
    \bar{S}_4&=\int d^4x\sqrt{-\bar{g}}\,\Omega^{-2}\left\{\frac12\,(\Mpl^2+\beta_1\sigma^2)\,\Omega\left[ \bar{R} +3 \bar\Box\ln\Omega-\frac32\,\bar{g}^{\mu\nu} \partial_\mu\ln\Omega\, \partial_\nu\ln\Omega  \right] 
    \right.\\
    &\left.+\sigma\sqrt{-\Omega\bar{g}^{\mu\nu}\partial_{\mu}\varphi\partial_\nu\varphi}\nonumber- \Mpl^2\Lambda-\tfrac{1}{2}\varphi^2-\tfrac{1}{2}m^2\sigma^2
\right\}\nonumber\\
&=\int d^4x\sqrt{-\bar{g}}\left\{\frac{\Mpl^2}2\,\left[ \bar{R} -\frac32\,\bar{g}^{\mu\nu} \partial_\mu\ln\Omega\, \partial_\nu\ln\Omega  \right] 
    +\Omega^{-3/2}\sigma\sqrt{-\bar{g}^{\mu\nu}\partial_{\mu}\varphi\partial_\nu\varphi}\nonumber\right.\\
    &-\left. \Omega^{-2}\left[\Mpl^2\Lambda+\tfrac{1}{2}\varphi^2+\tfrac{1}{2}m^2\sigma^2\right]\right\},
\label{eq:action_int_cusc_EF}
\end{align}
where we have neglected a total derivative term. If we further make the following field redefinition
\begin{equation}
    \Omega=e^{\alpha/\Mpl}\,,\qquad\sigma=\pm\Mpl\sqrt{\frac{e^{\alpha/\Mpl}-1}{\beta_1}}\,,
\end{equation}
so that $\alpha\beta_1>0$ (for $\sigma$ to be real), we finally find

\begin{align}
    {\bar S}_5^{\pm}&=\int d^4x\sqrt{-\bar{g}}\left\{\frac{\Mpl^2}2 \bar{R}-\frac{3}{4}\,\bar{g}^{\mu\nu} \partial_\mu\alpha\, \partial_\nu\alpha 
    -\Mpl V_1(\alpha)\sqrt{-\bar{g}^{\mu\nu}\partial_{\mu}\varphi\partial_\nu\varphi}- \Mpl^2\, V_2(\alpha,\varphi)\right\},
\label{eq:action_int_cusc_EF_alpha}
\end{align}
where the field $\varphi$ has dimensions of a squared mass, and we have defined
\begin{align}
V_1(\alpha)&=\mp e^{-3\alpha/2\Mpl}\sqrt{\frac{e^{\alpha/\Mpl}-1}{\beta_1}} \,,\\
V_2(\alpha,\varphi)&=e^{-2\alpha/\Mpl}\left[\Lambda+\tfrac{1}{2}\,\frac{\varphi^2}{\Mpl^2}+\frac{m^2}{2\beta_1}(e^{\alpha/\Mpl}-1)\right].
\end{align}
Notice that, since $\sigma\neq0$, also $\alpha\neq0$, but $\beta_1\alpha>0$ (the condition for $\sigma$ to be real, as explained above).

\section{{The perturbations in the Einstein frame}}\label{CP5}

The above action shows that a cuscuton field $\varphi$ is coupled with a canonical scalar field $\alpha$. Since the cuscuton, in unitary gauge, $\delta\varphi=0$, gives an effective time-dependent potential, as $\varphi=\varphi(t)$, then we may expect to find unity speed of propagation for the perturbative fields. This would match the results for the branch $A_1=0$ discussed already before. In any case, we will here show that this is indeed the case.

As usual, we consider the metric perturbations as given in Eq.\ \eqref{mBT1pert}. For this action, we need also to perturb the fields $\varphi$ and $\alpha$, as
\begin{equation}
    \varphi=\varphi(t)+\delta\varphi\,,\qquad
    \alpha = \alpha(t) + \delta\alpha\,.
\end{equation}
As for the background, we will solve the equations of motion for the lapse, the two scale factors $a$ and $b$, the scalar $\varphi(t)$, and $\sigma(t)$ for the following quantities: $V_2$, $\ddot{a}$, $\ddot{b}$, $\ddot\alpha$, and $V_1$. We are now ready to study the perturbation dynamics. 

\subsection{The perturbations on an FLRW manifold}

We first discuss the above action perturbatively on an FLRW manifold. We consider the same fields for the perturbation of the metric field as given in Eq.\ \eqref{eq:pert_metr_A}, and we set the gauge so that $\delta\varphi=0$, and $E=0$. For this FLRW manifold, since $b(t)=a(t)$, we do not have an equation determining $\ddot{b}$, so we can use the equations of motion for the background to set $V_2,\ddot{a},\ddot{\alpha}$, and $V_1$.

In the scalar sector, the metric perturbation $B$ sets an algebraic constraint, in Fourier space, for $\phi$ as a function of $\dot\psi$ and $\delta\alpha$. After that, on making the field redefinition $\psi=\psi_2+\dot{a}\delta\alpha/(a\dot\alpha)$, we can integrate out $\delta\alpha$ using its equation of motion, leaving only one field, $\psi_2$, in the reduced Lagrangian. Its kinetic term, in the high-$k$ regime, is $3a^5\dot{\alpha}^2/(4\dot{a}^2)$ and always satisfied. The speed of propagation for this mode is unity.

As for the other modes, only the tensor sector adds two degrees of freedom, which propagate luminally with a trivially satisfied no-ghost condition.

\subsection{Bianchi I: even modes}
Since the field $\varphi$ is bound to be timelike, it is natural and possible to choose the following gauge
\begin{equation}
    \delta\varphi=0\,,\quad
    E=0\,,\quad
\mu = 0\,.    
\end{equation}
Once more, the definition of the fields $E$ and $\mu$ above, together with the metric field perturbations, is given in Eq.\ \eqref{mBT1pert}.

After using a Fourier decomposition for all the fields, we first integrate out the field $B$. At this point, the field $\omega$, coming only linearly and without time-derivatives (after performing several integrations by parts), sets a constraint for the other fields. To simplify such a constraint, we perform the following field redefinition
\begin{equation}
    \psi=\bar\psi -\frac{2(q^2a^2+2k^2b^2)}{a^2q^2}\,\tau\,.
\end{equation}
Therefore, we can solve the $\omega$-constraint for the field $\phi$ in terms of the other fields. The action at this moment only depends on the fields $\bar\psi$, $\tau$, and $\delta\alpha$. The kinetic matrix of these fields has a vanishing determinant; therefore, we perform the following field redefinition
\begin{align}
    \bar\psi&=\tilde\psi +\frac{2(a^2\dot{b}q^2+ab\dot{a}q^2+2b^2\dot{b}k^2)}{ba^2q^2\dot{\alpha}}\,\delta\alpha\,,\\
    \tau&=\tilde\tau +\frac{\dot{b}}{b\dot{\alpha}}\,\delta\alpha\,.
\end{align}

At this point, the field $\delta\alpha$ becomes a Lagrange multiplier and we can integrate it out, leaving a reduced Lagrangian that is a function only of $\tilde\psi$ and $\tilde\tau$.

The no-ghost conditions for this 2-fields system are always satisfied both in the high $k$ and in the high $q$ regimes. They are proportional (up to positive background quantities) to the following two quantities
\begin{equation}
    \frac{2\Mpl^2\dot{b}^2+3b^2\dot{\alpha}^2}{b^2\dot{b}^2}>0\,,\qquad
    \frac{\dot{\alpha}^2 a^6b^2}{\dot{b}^2}>0\,.
\end{equation}
This shows that $\dot\alpha$ should not vanish on the dynamics.

The two squared speeds of propagation for the even modes are equal to unity both in the high-$k$ and high-$q$ limits. Therefore, this system exhibits no unstable behavior. This result is compatible with the ones shown for the branch $A_1=0$, the branch which can be connected to the interacting cuscuton action.

\subsection{Odd modes}

For the odd modes we choose the gauge $h=0$, (see Eq.\ \eqref{mBT1pert} for its definition together with that of the remaining odd modes). In this case, the only two odd modes variables consist of the variables $v$ and $\lambda$. Since $\alpha$ and $\varphi$ are scalars, they do not contribute to the odd modes. The field $v$ can be integrated out, leaving only the field $\lambda$ to propagate. Its no-ghost condition is trivially satisfied, and its squared speed of propagation is unity in both high-$k$ and high-$q$ regimes.

\section{Conclusion and Summary}\label{CP6}

The theory of three-form fields is a very peculiar theory: When minimally coupled to gravity, it propagates no (pseudo)scalar modes; however, upon introducing a mass term, a single scalar degree of freedom emerges. One might therefore naturally expect that introducing further non-minimal couplings would always excite such a scalar mode, in addition to the two tensor modes from the Einstein–Hilbert action, since the Ricci scalar effectively contributes as a mass term. Yet, at the same time, one should be careful on concluding if these three are the only degrees of freedom, and whether some additional pathologies emerge. 

Recently, it was shown in \cite{DeFelice:2025ykh} that if one considers a Proca theory with non-minimal coupling to gravity, not only does an additional mode appear, but its propagation speed vanishes, signaling strong coupling. Since the dual of a three-form field is precisely a vector field, this suggests that a similar pathology could arise in the three-form case as well.
The goal of this paper was to fill this gap by studying several aspects of the non-minimal 3-form theory in homogeneous, isotropic, and anisotropic universes. 

Overall, we have found that in the presence of non-minimal coupling with the Ricci scalar, the theory propagates three degrees of freedom: one scalar mode arising from the vector, dual to the 3-form due to the appearance of an effective mass, and two tensor modes, which are due to the gravity, in agreement with the previous claims in the literature. This is a positive result, as it signals that there exists another theory of vector fields, with an absent Maxwell kinetic term, that allows for non-minimal coupling to gravity, while at the same time preserving the number of modes for the non-trivial background solutions. 

First, we have studied the theory coupled to both the Ricci scalar and the Ricci tensor in the homogeneous and isotropic Universe. We have allowed for the most general case, including the non-vanishing background of the temporal component of the vector field\footnote{The non-vanishing spatial component would violate isotropy in the FLRW background.}. At the level of perturbations around this background, we have confirmed that the theory describes a scalar mode and two tensor modes. However, in contrast to the minimally coupled case, the two satisfy non-trivial no-ghost conditions. Moreover, due to the presence of the Ricci tensor, their speed of propagation is distorted -- it can be different from unity, and becomes strictly one if one sets the Ricci tensor to zero. 

In contrast to the homogeneous and isotropic Universe, the anisotropic Bianchi Type 1 allows for a more general solution to the vector field, in the non-vanishing spatial component. For simplicity, we have set the Ricci tensor to zero and focused on the coupling with the Ricci scalar only. Curiously, this theory shows similar behavior to the case of non-minimal Proca, in the sense that it allows for two different branches -- one with vanishing $A_1$ component, and one with a non-vanishing value. Both branches preserve the number of degrees of freedom, giving rise to two even modes and one odd one. However, the behavior of the modes significantly differs between the two. 

In the $A_1 \neq 0$ branch, we obtain the no-ghost conditions, which are slightly modified relative to the isotropic case due to the presence of a non-vanishing background spatial component, both in the high-momentum limit along the anisotropic direction and within the homogeneous subspace. At high momenta—both along the anisotropic direction and within the homogeneous subspace—the odd mode propagates with unit speed. The even sector, however, displays a more intriguing behavior: the speed of propagation is unity in the homogeneous subspace, but, in the anisotropic direction, the speed of propagation for one mode vanishes, whereas for the other is constant, and its group velocity vanishes. This indicates a strong coupling, similarly to the Proca case, though importantly, without the appearance of additional degrees of freedom.

In contrast to the $A_1\neq 0$ branch, $A_1=0$ contains modes with non-trivial no-ghost conditions, but unit speed of propagation in both high-k and high-q limits. Initially, one might find the analysis for this branch very non-trivial. In the former case, one finds a mode that naturally drops out from the action ($\pi$), which is not the case if $A_1=0$. However, this branch is, on the other hand, special, as it allows for a reformulation that clarifies the count of the degrees of freedom. In particular, we have shown that it can be written as a theory of a cuscuton field, coupled to a constrained scalar with non-minimal coupling to gravity. This makes the study of the dof much simpler: it is well known that the cuscuton does not give rise to new dof. Therefore, the scalar mode emerges from the constrained scalar field. Notably, this formulation provides non-trivial no-ghost conditions that are possible to satisfy, and modes with unit speed of propagation, most conveniently shown in the corresponding Einstein frame. 

Our results can be summarized by the Figure 1. 
\begin{figure}[h!]
    \centering

        \centering
        \includegraphics[width=\linewidth]{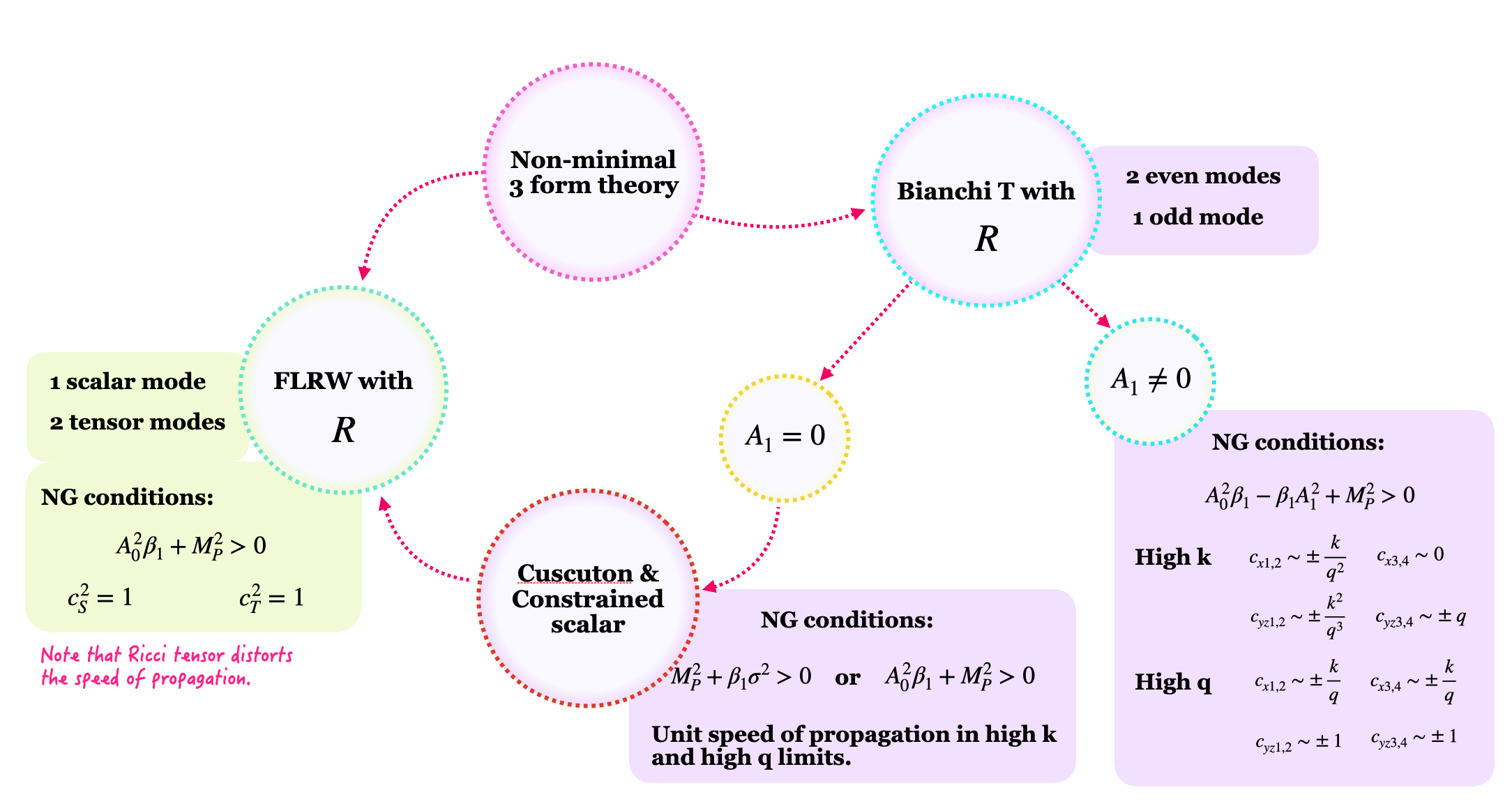}

    \caption{The summary of the main results. }
   
\end{figure}

Given the recent developments in the theories of vector fields, one might think that if the non-minimal coupling is allowed, one can only have a combination of the Ricci scalar and Ricci tensor that forms the Einstein tensor to have a well-behaved theory. Our analysis shows that the vector field, dual to the 3-form, could also be an interesting potential candidate, especially given the existence of the $A_1= 0$ branch. 

However, at the same time, one should be careful about the $A_1\neq 0$ branch, which shows peculiar behavior in the high-k limit. \textcolor{Black}{Due to the vanishing speed of propagation along the x-direction in certain cases, there is a high chance that this branch might be strongly coupled. In particular, this appears for one of the modes in the high-k limit, while in the high-q limit, the speed of propagation vanishes for both of the modes, when $q\to\infty$.  }
This requires a dedicated analysis of the non-linear terms. However, if it takes place, it does not mean that the theory is necessarily unhealthy -- the strong coupling could potentially be resolved in a manner similar to the Vainshtein mechanism, applied originally to massive gravity and later other gauge theories as well \cite{Vainshtein:1972sx, Deffayet:2001uk, Gruzinov:2001hp, Dvali:2006su, Dvali:2007kt, Hell:2021wzm,  Hell:2021oea, Hell:2022wci, Hell:2023mph, Hell:2025uoc, Hell:2025pso}.  Whether this takes place or not is an exciting question that will be clarified in future works. 

\textcolor{Black}{The physical applications of the findings of this paper are particularly interesting. The 3-form field is usually associated with a massive axion -- one of the prominent dark matter candidates \cite{Quevedo:1996uu, Dvali:2005an}\footnote{See also \cite{Hell:2021oea}, where it was pointed out that the relation between the 3-form and scalar field could break-down for particular self-interactions, due to the strong coupling. }. However, here we show that in the presence of non-minimal coupling to the Ricci scalar, the 3-form can be connected to only one branch of solutions. This is not possible for the second branch, $A_1\neq0$, where one thus cannot connect the axion with the 3-form. Another important point is that the scalar field branch involves both a cuscuton and the constrained scalar, analyzed in \cite{Hell:2025lgn}. This is particularly intriguing, as the constrained scalar described in \cite{Hell:2025lgn} can admit dark energy evolution with phantom-like equation of state, while being theoretically healthy. This relation thus opens an intriguing avenue of exploring accelerating solutions also in the presence of cuscuton, and thus also potentially underpinning the late-time acceleration of the Universe. }

Overall, based on the analysis done in this work, it appears that the non-minimal theory of a 3-form is special: unlike other massive gauge theories, it preserves the degrees of freedom,  admits well-behaved branches, and also those that need further investigations into the non-linear regime, which is an exciting avenue to pursue in future research. 

\begin{center}
\textbf{\textsc{Acknowledgments}}
\end{center}
\textit{A. H. would like to thank Elisa G. M. Ferreira, and Misao Sasaki for very useful discussions. 
A. H. also thanks the organizers of Corfu2025;  MPP Munich, and, in particular, Prof. Dieter L\"ust, for hospitality during
her visit, when part of this work was carried out, and for useful comment on the paper. The work of A. H. was supported in part by JSPS KAKENHI No.~24K00624, and 
by the World Premier International Research Center Initiative (WPI), MEXT, Japan. }

\section*{Appendix}
In order to have a clearer view of the theory, in the following, we will present the coefficients for the non-minimal 3-form action in the homogeneous and isotropic expanding Universe, separating the coefficients for the scalar, vector, and tensor modes. We will also apply a similar procedure in the case of an anisotropic universe; however, due to the large number of terms, we will omit the expressions for the coefficients. 
\subsection*{ A.1 The scalar mode coefficients}\label{A1}
The following are the coefficients that correspond to the action (\ref{LagrdenFLRW}). 
\begin{equation*}
    \begin{split}
        c_1&=-3 a^{3} \left(\left(-\frac{3}{2}+\beta_{1}+\beta_{2}\right) A_{0}^{2}+\Mpl^{2}\right)\qquad 
        c_2=\frac{A_{0}^{2} a^{3}}{2}\qquad 
        c_3=\frac{a^{3}}{2}\qquad
        c_4=\frac{a^{3}}{2}\\
        c_5&=\left(-6 \beta_{1}-3 \beta_{2}+3\right) a^{3} A_{0}\qquad 
        c_6=\left(6 \beta_{1}+3 \beta_{2}-3\right) a^{3} A_{0}^{2}\qquad
        c_7=3 a^{2} \left(3 A_{0} a H +\dot{A}_0 a \right)\\
        c_8&=-3 a^{2} \left(3 A_{0} a H +\dot{A}_0 a \right)\qquad
        c_9=-6 \left(\frac{\dot{A}_0 a}{6}+a H A_{0} \left(\beta_{1}+\frac{\beta_{2}}{2}+1\right)\right) a^{2}\\
        c_{10}&=-9 a^{2} \left(\frac{A_{0}\dot{A}_0 a}{3}+\left(\left(\beta_{1}+\frac{\beta_{2}}{2}+1\right) A_{0}^{2}+\Mpl^{2}\right) a H \right)\\
        c_{11}&=-3 \left(3 \left(\beta_{1}-\frac{\beta_{2}}{2}+3\right) a H A_{0}^{2}-6 A_{0} \left(\beta_{1}+\frac{\beta_{2}}{2}-\frac{1}{2}\right)\dot{A}_0 a +a H \,\Mpl^{2}\right) a^{2}\\
        c_{12}&=-6 a^{2} \left(\left(\beta_{1}+\frac{\beta_{2}}{2}-\frac{1}{2}\right)\dot{A}_0 a -a H A_{0} \left(\beta_{1}-\frac{\beta_{2}}{2}+3\right)\right)\\
        c_{13}&=-a\qquad
        c_{14}=A_{0} a\qquad
        c_{15}=A_{0} \left(2 \beta_{2}-3\right) a\qquad
       c_{16}=\frac{1}{2 a}\qquad
       c_{17}=-3 a H\\
       c_{18}&=\left(-2 \beta_{1}-\beta_{2}\right) a A_{0}\qquad
        c_{19}=-4 a A_{0} \beta_{1}\\
        c_{20}&=\frac{1}{2 a}\left(-m^{2} a^{2}+\left(-6 a^{2} H^{2}+R \,a^{2}\right) \left(\beta_{1}+\frac{\beta_{2}}{6}\right)+6 a^{2} H^{2} \left(\beta_{1}+\frac{\beta_{2}}{3}\right)\right)\\
        c_{21}&=2\dot{A}_0 a -2 a H A_{0} \left(\beta_{2}-3\right)\qquad
        c_{22}=a \left(2 \beta_{1}+\beta_{2}\right) A_{0}^{2}\\
        c_{23}&=-2 a \left(-A_{0}^{2} \beta_{1}+\Mpl^{2}\right)\qquad
        c_{24}=-a \left(A_{0}^{2} \beta_{1}+\Mpl^{2}\right)\\
        c_{25}&=-\frac{a }{2}\left(m^{2} a^{2}-\left(-6 a^{2} H^{2}+R \,a^{2}\right) \left(\beta_{1}+\frac{\beta_{2}}{2}-\frac{1}{2}\right)-6 \left(\beta_{1}+\frac{1}{2}\right) a^{2} H^{2}\right)
    \end{split}
\end{equation*}

\begin{equation*}
    \begin{split}
    c_{26}&=-\frac{3}{4} \left(\left(A_{0}^{2} m^{2}-4 A_{0} \left(\beta_{1}+\frac{\beta_{2}}{2}-\frac{1}{2}\right)\ddot{A}_0+\left(-4 \beta_{1}-2 \beta_{2}+1\right)\dot{A}_0^{2}+2 \Mpl^{2} \Lambda \right) a^{2}\right.\\&\left.-\frac{2}{3 } \left(-\frac{3}{2}+\beta_{1}+\beta_{2}\right) \left(-6 a^{2} H^{2}+R \,a^{2}\right) A_{0}^{2}-8  H \left(-\frac{3}{2}+\beta_{1}+\beta_{2}\right)\dot{A}_0 A_{0}-\frac{2}{3 } \left(-6 a^{2} H^{2}+R \,a^{2}\right) \Mpl^{2}\right.\\&\left.  -2 \left(\left(-\frac{3}{2}+\beta_{1}+\beta_{2}\right) A_{0}^{2}+\Mpl^{2}\right) a^{2} H^{2}\right) a\\
            c_{27}&=-\frac{3 a }{4}\left(\left(m^{2} a^{2}-\frac{10 \left(-6 a^{2} H^{2}+R \,a^{2}\right) \left(\beta_{1}+\frac{\beta_{2}}{2}-\frac{1}{2}\right)}{3}-10 a^{2} H^{2} \left(\beta_{1}-\beta_{2}+\frac{5}{2}\right)\right) A_{0}^{2}\right.\\&\left.+20 a \left(\frac{\ddot{A}_0 a}{6}+\dot{A}_0 a H \left(\beta_{1}+\frac{\beta_{2}}{2}\right)\right) A_{0}+\frac{\left(-2 \Mpl^{2} \Lambda -5\dot{A}_0^{2}\right) a^{2}}{3}+6 \Mpl^{2} a^{2} H^{2}\right)\\
             c_{28}&=\frac{3 a }{2}\left(\left(A_{0}^{2} m^{2}-2 \Mpl^{2} \Lambda -3\dot{A}_0^{2}\right) a^{2}-6 a^{2} H^{2} \left(\left(3 \beta_{1}+\frac{9}{2}\right) A_{0}^{2}+\Mpl^{2}\right)\right.\\&\left.+\left(-\frac{3 \left(\beta_{1}+\frac{\beta_{2}}{2}\right) \left(-6 a^{2} H^{2}+R \,a^{2}\right) A_{0}^{2}}{a}-18 A_{0} a H\dot{A}_0-\frac{\left(-6 a^{2} H^{2}+R \,a^{2}\right) \Mpl^{2}}{a}\right) a \right)\\
         c_{29}&=-3 a \left(\left(m^{2} a^{2}-\left(\beta_{1}+\frac{\beta_{2}}{2}\right) \left(-6 a^{2} H^{2}+R \,a^{2}\right)-6 a^{2} H^{2} \left(\beta_{1}+\frac{3}{2}\right)\right) A_{0}-3 a^{2} H\dot{A}_0\right)\\
        c_{30}&=\left(\left(m^{2} a^{2}-3 \left(\beta_{1}+\frac{\beta_{2}}{2}\right) \left(-6 a^{2} H^{2}+R \,a^{2}\right)-18 a^{2} H^{2} \left(\beta_{1}+\frac{3}{2}\right)\right) A_{0}-9 a^{2} H\dot{A}_0\right) a
    \end{split}
\end{equation*}
\subsection*{ A.2 The vector mode coefficients}\label{A2}
The following are the coefficients for the vector modes in the FLRW Universe that correspond to the action (\ref{actVec}). 
\begin{equation*}
    \begin{split}
        d_1&=-a^{2} \left(3 A_{0} a H +\dot{A}_0 a \right) \qquad
        d_2=-\frac{a \left(\left(\beta_{1}+\beta_{2}\right) A_{0}^{2}+\Mpl^{2}\right)}{4}\qquad
        d_3=-\frac{a A_{0} \beta_{2}}{2}\\
        d_4&=a \left(\left(m^{2} a^{2}-\left(\beta_{1}+\frac{\beta_{2}}{2}\right) \left(-6 a^{2} H^{2}+R \,a^{2}\right)-6 a^{2} H^{2} \left(\beta_{1}+\frac{3}{2}\right)\right) A_{0}-3 a^{2} H \dot{A}_0\right)\\
        d_5&=\frac{a}{4} \left(\left(m^{2} a^{2}-2 \left(\beta_{1}+\frac{\beta_{2}}{2}-\frac{1}{2}\right) \left(-6 a^{2} H^{2}+R \,a^{2}\right)-6 a^{2} H^{2} \left(\beta_{1}-\beta_{2}+\frac{5}{2}\right)\right) A_{0}^{2}\right.\\&\left.+12 a \left(\frac{\ddot{A}_0 a}{6}+a H \dot{A}_0 \left(\beta_{1}+\frac{\beta_{2}}{2}\right)\right) A_{0}+\left(-2 \Lambda \,\Mpl^{2}-\dot{A}_0^{2}\right) a^{2}+6 \Mpl^{2} a^{2} H^{2}\right)\\
        d_6&=\frac{a }{2}\left(m^{2} a^{2}-\left(\beta_{1}+\frac{\beta_{2}}{6}\right) \left(-6 a^{2} H^{2}+R \,a^{2}\right)-6 a^{2} H^{2} \left(\beta_{1}+\frac{\beta_{2}}{3}\right)\right)
    \end{split}
\end{equation*}

\subsection*{ A.3 The tensor mode coefficients}\label{A3}
The following are the coefficients for the tensor modes, arising in the action (\ref{actTen}). 
\begin{equation*}
    \begin{split}
        e_1=\frac{a^{3}}{8} \left(\left(\beta_{1}+\beta_{2}\right) A_{0}^{2}+\Mpl^{2}\right)\qquad
   e_2=\frac{a \left(A_{0}^{2} \beta_{1}+\Mpl^{2}\right)}{8}
    \end{split}
\end{equation*}

\begin{equation*}
   \begin{split}
   e_3&=\frac{a}{8} \left(\left(m^{2} A_{0}^{2}-4 A_{0} \left(\beta_{1}+\frac{\beta_{2}}{2}-\frac{1}{2}\right) \ddot{A}_0+\left(-4 \beta_{1}-2 \beta_{2}+1\right) \dot{A}_0^{2}+2 \Lambda \,\Mpl^{2}\right) a^{2}\right.\\&\left.+\left(-\frac{2 \left(-\frac{3}{2}+\beta_{1}+\beta_{2}\right) \left(-6 a^{2} H^{2}+R \,a^{2}\right) A_{0}^{2}}{3 a}-8 \left(-\frac{3}{2}+\beta_{1}+\beta_{2}\right) \dot{A}_0 a H A_{0}\right.\right.\\&\left.\left.-\frac{2 \left(-6 a^{2} H^{2}+R \,a^{2}\right) \Mpl^{2}}{3 a}\right) a -2 \left(\left(-\frac{3}{2}+\beta_{1}+\beta_{2}\right) A_{0}^{2}+\Mpl^{2}\right) a^{2} H^{2}\right)
   \end{split}
\end{equation*}

\bibliographystyle{utphys}
\bibliography{3formpaper}{}

\providecommand{\href}[2]{#2}\begingroup\raggedright\begin{thebibliography}{100}

\bibitem{Gates:1980ay}
S.~J. Gates, Jr., ``{SUPER P FORM GAUGE SUPERFIELDS},'' \href{http://dx.doi.org/10.1016/0550-3213(81)90225-X}{{\em Nucl. Phys. B} {\bf 184} (1981)  381--390}.

\bibitem{Gates:1980az}
S.~J. Gates, Jr. and W.~Siegel, ``{VARIANT SUPERFIELD REPRESENTATIONS},'' \href{http://dx.doi.org/10.1016/0550-3213(81)90281-9}{{\em Nucl. Phys. B} {\bf 187} (1981)  389--396}.

\bibitem{Binetruy:1996xw}
P.~Binetruy, F.~Pillon, G.~Girardi, and R.~Grimm, ``{The Three form multiplet in supergravity},'' \href{http://dx.doi.org/10.1016/0550-3213(96)00370-7}{{\em Nucl. Phys. B} {\bf 477} (1996)  175--202}, \href{http://arxiv.org/abs/hep-th/9603181}{{\tt arXiv:hep-th/9603181}}.

\bibitem{Ovrut:1997ur}
B.~A. Ovrut and D.~Waldram, ``{Membranes and three form supergravity},'' \href{http://dx.doi.org/10.1016/S0550-3213(97)00510-5}{{\em Nucl. Phys. B} {\bf 506} (1997)  236--266}, \href{http://arxiv.org/abs/hep-th/9704045}{{\tt arXiv:hep-th/9704045}}.

\bibitem{Bousso:2000xa}
R.~Bousso and J.~Polchinski, ``{Quantization of four form fluxes and dynamical neutralization of the cosmological constant},'' \href{http://dx.doi.org/10.1088/1126-6708/2000/06/006}{{\em JHEP} {\bf 06} (2000)  006}, \href{http://arxiv.org/abs/hep-th/0004134}{{\tt arXiv:hep-th/0004134}}.

\bibitem{Feng:2000if}
J.~L. Feng, J.~March-Russell, S.~Sethi, and F.~Wilczek, ``{Saltatory relaxation of the cosmological constant},'' \href{http://dx.doi.org/10.1016/S0550-3213(01)00097-9}{{\em Nucl. Phys. B} {\bf 602} (2001)  307--328}, \href{http://arxiv.org/abs/hep-th/0005276}{{\tt arXiv:hep-th/0005276}}.

\bibitem{Nishino:2009zz}
H.~Nishino and S.~Rajpoot, ``{Alternative auxiliary fields for chiral multiplets},'' \href{http://dx.doi.org/10.1103/PhysRevD.80.127701}{{\em Phys. Rev. D} {\bf 80} (2009)  127701}.

\bibitem{Bandos:2010yy}
I.~A. Bandos and C.~Meliveo, ``{Superfield equations for the interacting system of D=4 N=1 supermembrane and scalar multiplet},'' \href{http://dx.doi.org/10.1016/j.nuclphysb.2011.03.010}{{\em Nucl. Phys. B} {\bf 849} (2011)  1--27}, \href{http://arxiv.org/abs/1011.1818}{{\tt arXiv:1011.1818 [hep-th]}}.

\bibitem{Bandos:2011fw}
I.~A. Bandos and C.~Meliveo, ``{Three form potential in (special) minimal supergravity superspace and supermembrane supercurrent},'' \href{http://dx.doi.org/10.1088/1742-6596/343/1/012012}{{\em J. Phys. Conf. Ser.} {\bf 343} (2012)  012012}, \href{http://arxiv.org/abs/1107.3232}{{\tt arXiv:1107.3232 [hep-th]}}.

\bibitem{Groh:2012tf}
K.~Groh, J.~Louis, and J.~Sommerfeld, ``{Duality and Couplings of 3-Form-Multiplets in N=1 Supersymmetry},'' \href{http://dx.doi.org/10.1007/JHEP05(2013)001}{{\em JHEP} {\bf 05} (2013)  001}, \href{http://arxiv.org/abs/1212.4639}{{\tt arXiv:1212.4639 [hep-th]}}.

\bibitem{Nishino:2013oea}
H.~Nishino and S.~Rajpoot, ``{$N = 1$ supersymmetric Proca{\textendash}Stueckelberg mechanism for extra vector multiplet},'' \href{http://dx.doi.org/10.1016/j.nuclphysb.2014.08.003}{{\em Nucl. Phys. B} {\bf 887} (2014)  265--275}, \href{http://arxiv.org/abs/1309.6393}{{\tt arXiv:1309.6393 [hep-th]}}.

\bibitem{Bielleman:2015ina}
S.~Bielleman, L.~E. Ibanez, and I.~Valenzuela, ``{Minkowski 3-forms, Flux String Vacua, Axion Stability and Naturalness},'' \href{http://dx.doi.org/10.1007/JHEP12(2015)119}{{\em JHEP} {\bf 12} (2015)  119}, \href{http://arxiv.org/abs/1507.06793}{{\tt arXiv:1507.06793 [hep-th]}}.

\bibitem{Morais:2016bev}
J.~Morais, M.~Bouhmadi-L{\'o}pez, K.~Sravan~Kumar, J.~Marto, and Y.~Tavakoli, ``{Interacting 3-form dark energy models: distinguishing interactions and avoiding the Little Sibling of the Big Rip},'' \href{http://dx.doi.org/10.1016/j.dark.2016.11.002}{{\em Phys. Dark Univ.} {\bf 15} (2017)  7--30}, \href{http://arxiv.org/abs/1608.01679}{{\tt arXiv:1608.01679 [gr-qc]}}.

\bibitem{Bouhmadi-Lopez:2016dzw}
M.~Bouhmadi-L{\'o}pez, J.~Marto, J.~Morais, and C.~M. Silva, ``{Cosmic infinity: A dynamical system approach},'' \href{http://dx.doi.org/10.1088/1475-7516/2017/03/042}{{\em JCAP} {\bf 03} (2017)  042}, \href{http://arxiv.org/abs/1611.03100}{{\tt arXiv:1611.03100 [gr-qc]}}.

\bibitem{Buchbinder:2017vnb}
E.~I. Buchbinder and S.~M. Kuzenko, ``{Three-form multiplet and supersymmetry breaking},'' \href{http://dx.doi.org/10.1007/JHEP09(2017)089}{{\em JHEP} {\bf 09} (2017)  089}, \href{http://arxiv.org/abs/1705.07700}{{\tt arXiv:1705.07700 [hep-th]}}.

\bibitem{Farakos:2016hly}
F.~Farakos, A.~Kehagias, D.~Racco, and A.~Riotto, ``{Scanning of the Supersymmetry Breaking Scale and the Gravitino Mass in Supergravity},'' \href{http://dx.doi.org/10.1007/JHEP06(2016)120}{{\em JHEP} {\bf 06} (2016)  120}, \href{http://arxiv.org/abs/1605.07631}{{\tt arXiv:1605.07631 [hep-th]}}.

\bibitem{Bandos:2016xyu}
I.~Bandos, M.~Heller, S.~M. Kuzenko, L.~Martucci, and D.~Sorokin, ``{The Goldstino brane, the constrained superfields and matter in $ \mathcal{N}=1 $ supergravity},'' \href{http://dx.doi.org/10.1007/JHEP11(2016)109}{{\em JHEP} {\bf 11} (2016)  109}, \href{http://arxiv.org/abs/1608.05908}{{\tt arXiv:1608.05908 [hep-th]}}.

\bibitem{Aoki:2016rfz}
S.~Aoki, T.~Higaki, Y.~Yamada, and R.~Yokokura, ``{Abelian tensor hierarchy in 4D ${\cal N}=1$ conformal supergravity},'' \href{http://dx.doi.org/10.1007/JHEP09(2016)148}{{\em JHEP} {\bf 09} (2016)  148}, \href{http://arxiv.org/abs/1606.04448}{{\tt arXiv:1606.04448 [hep-th]}}.

\bibitem{Carta:2016ynn}
F.~Carta, F.~Marchesano, W.~Staessens, and G.~Zoccarato, ``{Open string multi-branched and K{\"a}hler potentials},'' \href{http://dx.doi.org/10.1007/JHEP09(2016)062}{{\em JHEP} {\bf 09} (2016)  062}, \href{http://arxiv.org/abs/1606.00508}{{\tt arXiv:1606.00508 [hep-th]}}.

\bibitem{Valenzuela:2016yny}
I.~Valenzuela, ``{Backreaction Issues in Axion Monodromy and Minkowski 4-forms},'' \href{http://dx.doi.org/10.1007/JHEP06(2017)098}{{\em JHEP} {\bf 06} (2017)  098}, \href{http://arxiv.org/abs/1611.00394}{{\tt arXiv:1611.00394 [hep-th]}}.

\bibitem{Farakos:2017jme}
F.~Farakos, S.~Lanza, L.~Martucci, and D.~Sorokin, ``{Three-forms in Supergravity and Flux Compactifications},'' \href{http://dx.doi.org/10.1140/epjc/s10052-017-5185-y}{{\em Eur. Phys. J. C} {\bf 77} (2017) no.~9, 602}, \href{http://arxiv.org/abs/1706.09422}{{\tt arXiv:1706.09422 [hep-th]}}.

\bibitem{Farakos:2017ocw}
F.~Farakos, S.~Lanza, L.~Martucci, and D.~Sorokin, ``{Three-forms, Supersymmetry and String Compactifications},'' \href{http://dx.doi.org/10.1134/S1063779618050192}{{\em Phys. Part. Nucl.} {\bf 49} (2018) no.~5, 823--828}, \href{http://arxiv.org/abs/1712.09366}{{\tt arXiv:1712.09366 [hep-th]}}.

\bibitem{Kuzenko:2017vil}
S.~M. Kuzenko and G.~Tartaglino-Mazzucchelli, ``{Complex three-form supergravity and membranes},'' \href{http://dx.doi.org/10.1007/JHEP12(2017)005}{{\em JHEP} {\bf 12} (2017)  005}, \href{http://arxiv.org/abs/1710.00535}{{\tt arXiv:1710.00535 [hep-th]}}.

\bibitem{Bandos:2018gjp}
I.~Bandos, F.~Farakos, S.~Lanza, L.~Martucci, and D.~Sorokin, ``{Three-forms, dualities and membranes in four-dimensional supergravity},'' \href{http://dx.doi.org/10.1007/JHEP07(2018)028}{{\em JHEP} {\bf 07} (2018)  028}, \href{http://arxiv.org/abs/1803.01405}{{\tt arXiv:1803.01405 [hep-th]}}.

\bibitem{Becker:2017zwe}
K.~Becker, M.~Becker, D.~Butter, S.~Guha, W.~D. Linch, and D.~Robbins, ``{Eleven-dimensional supergravity in 4D, $N = 1$ superspace},'' \href{http://dx.doi.org/10.1007/JHEP11(2017)199}{{\em JHEP} {\bf 11} (2017)  199}, \href{http://arxiv.org/abs/1709.07024}{{\tt arXiv:1709.07024 [hep-th]}}.

\bibitem{Herraez:2018vae}
A.~Herraez, L.~E. Ibanez, F.~Marchesano, and G.~Zoccarato, ``{The Type IIA Flux Potential, 4-forms and Freed-Witten anomalies},'' \href{http://dx.doi.org/10.1007/JHEP09(2018)018}{{\em JHEP} {\bf 09} (2018)  018}, \href{http://arxiv.org/abs/1802.05771}{{\tt arXiv:1802.05771 [hep-th]}}.

\bibitem{Cribiori:2018jjh}
N.~Cribiori and S.~Lanza, ``{On the dynamical origin of parameters in $\mathcal {N}=2$ supersymmetry},'' \href{http://dx.doi.org/10.1140/epjc/s10052-019-6545-6}{{\em Eur. Phys. J. C} {\bf 79} (2019) no.~1, 32}, \href{http://arxiv.org/abs/1810.11425}{{\tt arXiv:1810.11425 [hep-th]}}.

\bibitem{Nitta:2018vyc}
M.~Nitta and R.~Yokokura, ``{Topological couplings in higher derivative extensions of supersymmetric three-form gauge theories},'' \href{http://dx.doi.org/10.1007/JHEP05(2019)102}{{\em JHEP} {\bf 05} (2019)  102}, \href{http://arxiv.org/abs/1810.12678}{{\tt arXiv:1810.12678 [hep-th]}}.

\bibitem{Nitta:2018yzb}
M.~Nitta and R.~Yokokura, ``{Higher derivative three-form gauge theories and their supersymmetric extension},'' \href{http://dx.doi.org/10.1007/JHEP10(2018)146}{{\em JHEP} {\bf 10} (2018)  146}, \href{http://arxiv.org/abs/1809.03957}{{\tt arXiv:1809.03957 [hep-th]}}.

\bibitem{Bandos:2019qok}
I.~Bandos, S.~Lanza, and D.~Sorokin, ``{Supermembranes and domain walls in $\mathcal N=1$, $D=4$ SYM},'' \href{http://dx.doi.org/10.1007/JHEP12(2019)021}{{\em JHEP} {\bf 12} (2019)  021}, \href{http://arxiv.org/abs/1905.02743}{{\tt arXiv:1905.02743 [hep-th]}}. [Erratum: JHEP 05, 031 (2020)].

\bibitem{Lanza:2019xxg}
S.~Lanza, F.~Marchesano, L.~Martucci, and D.~Sorokin, ``{How many fluxes fit in an EFT?},'' \href{http://dx.doi.org/10.1007/JHEP10(2019)110}{{\em JHEP} {\bf 10} (2019)  110}, \href{http://arxiv.org/abs/1907.11256}{{\tt arXiv:1907.11256 [hep-th]}}.

\bibitem{Bandos:2019lps}
I.~Bandos, ``{Superstring at the boundary of open supermembrane interacting with D=4 supergravity and matter supermultiplets},'' \href{http://dx.doi.org/10.1007/JHEP12(2019)106}{{\em JHEP} {\bf 12} (2019)  106}, \href{http://arxiv.org/abs/1906.09872}{{\tt arXiv:1906.09872 [hep-th]}}.

\bibitem{Dudas:2019gxd}
E.~Dudas, P.~Lamba, and S.~K. Vempati, ``{Diluting SUSY flavour problem on the Landscape},'' \href{http://dx.doi.org/10.1016/j.physletb.2020.135404}{{\em Phys. Lett. B} {\bf 804} (2020)  135404}, \href{http://arxiv.org/abs/1912.12839}{{\tt arXiv:1912.12839 [hep-ph]}}.

\bibitem{Cribiori:2020wch}
N.~Cribiori, F.~Farakos, and G.~Tringas, ``{Three-forms and Fayet-Iliopoulos terms in Supergravity: Scanning Planck mass and BPS domain walls},'' \href{http://dx.doi.org/10.1007/JHEP05(2020)060}{{\em JHEP} {\bf 05} (2020)  060}, \href{http://arxiv.org/abs/2001.05757}{{\tt arXiv:2001.05757 [hep-th]}}.

\bibitem{Hawking:1984hk}
S.~W. Hawking, ``{The Cosmological Constant Is Probably Zero},'' \href{http://dx.doi.org/10.1016/0370-2693(84)91370-4}{{\em Phys. Lett. B} {\bf 134} (1984)  403}.

\bibitem{Brown:1987dd}
J.~D. Brown and C.~Teitelboim, ``{Dynamical Neutralization of the Cosmological Constant},'' \href{http://dx.doi.org/10.1016/0370-2693(87)91190-7}{{\em Phys. Lett. B} {\bf 195} (1987)  177--182}.

\bibitem{Turok:1998he}
N.~Turok and S.~W. Hawking, ``{Open inflation, the four form and the cosmological constant},'' \href{http://dx.doi.org/10.1016/S0370-2693(98)00651-0}{{\em Phys. Lett. B} {\bf 432} (1998)  271--278}, \href{http://arxiv.org/abs/hep-th/9803156}{{\tt arXiv:hep-th/9803156}}.

\bibitem{Kaloper:2022jpv}
N.~Kaloper and A.~Westphal, ``{Quantum-mechanical mechanism for reducing the cosmological constant},'' \href{http://dx.doi.org/10.1103/PhysRevD.106.L101701}{{\em Phys. Rev. D} {\bf 106} (2022) no.~10, L101701}, \href{http://arxiv.org/abs/2204.13124}{{\tt arXiv:2204.13124 [hep-th]}}.

\bibitem{Kaloper:2022oqv}
N.~Kaloper, ``{Hidden variables of gravity and geometry and the cosmological constant problem},'' \href{http://dx.doi.org/10.1103/PhysRevD.106.065009}{{\em Phys. Rev. D} {\bf 106} (2022) no.~6, 065009}, \href{http://arxiv.org/abs/2202.06977}{{\tt arXiv:2202.06977 [hep-th]}}.

\bibitem{Liu:2023vqp}
Y.~Liu, A.~Padilla, and F.~G. Pedro, ``{The cosmological constant is probably still zero},'' \href{http://dx.doi.org/10.1007/JHEP10(2023)014}{{\em JHEP} {\bf 10} (2023)  014}, \href{http://arxiv.org/abs/2303.17723}{{\tt arXiv:2303.17723 [hep-th]}}.

\bibitem{Koivisto:2009fb}
T.~S. Koivisto and N.~J. Nunes, ``{Inflation and dark energy from three-forms},'' \href{http://dx.doi.org/10.1103/PhysRevD.80.103509}{{\em Phys. Rev. D} {\bf 80} (2009)  103509}, \href{http://arxiv.org/abs/0908.0920}{{\tt arXiv:0908.0920 [astro-ph.CO]}}.

\bibitem{Koivisto:2009ew}
T.~S. Koivisto and N.~J. Nunes, ``{Three-form cosmology},'' \href{http://dx.doi.org/10.1016/j.physletb.2010.01.051}{{\em Phys. Lett. B} {\bf 685} (2010)  105--109}, \href{http://arxiv.org/abs/0907.3883}{{\tt arXiv:0907.3883 [astro-ph.CO]}}.

\bibitem{Kobayashi:2009hj}
T.~Kobayashi and S.~Yokoyama, ``{Gravitational waves from p-form inflation},'' \href{http://dx.doi.org/10.1088/1475-7516/2009/05/004}{{\em JCAP} {\bf 05} (2009)  004}, \href{http://arxiv.org/abs/0903.2769}{{\tt arXiv:0903.2769 [astro-ph.CO]}}.

\bibitem{Koivisto:2011rm}
T.~S. Koivisto and F.~R. Urban, ``{Three-magnetic fields},'' \href{http://dx.doi.org/10.1103/PhysRevD.85.083508}{{\em Phys. Rev. D} {\bf 85} (2012)  083508}, \href{http://arxiv.org/abs/1112.1356}{{\tt arXiv:1112.1356 [astro-ph.CO]}}.

\bibitem{DeFelice:2012jt}
A.~De~Felice, K.~Karwan, and P.~Wongjun, ``{Stability of the 3-form field during inflation},'' \href{http://dx.doi.org/10.1103/PhysRevD.85.123545}{{\em Phys. Rev. D} {\bf 85} (2012)  123545}, \href{http://arxiv.org/abs/1202.0896}{{\tt arXiv:1202.0896 [hep-ph]}}.

\bibitem{Mulryne:2012ax}
D.~J. Mulryne, J.~Noller, and N.~J. Nunes, ``{Three-form inflation and non-Gaussianity},'' \href{http://dx.doi.org/10.1088/1475-7516/2012/12/016}{{\em JCAP} {\bf 12} (2012)  016}, \href{http://arxiv.org/abs/1209.2156}{{\tt arXiv:1209.2156 [astro-ph.CO]}}.

\bibitem{Koivisto:2012xm}
T.~S. Koivisto and N.~J. Nunes, ``{Coupled three-form dark energy},'' \href{http://dx.doi.org/10.1103/PhysRevD.88.123512}{{\em Phys. Rev. D} {\bf 88} (2013) no.~12, 123512}, \href{http://arxiv.org/abs/1212.2541}{{\tt arXiv:1212.2541 [astro-ph.CO]}}.

\bibitem{Kumar:2014oka}
K.~S. Kumar, J.~Marto, N.~J. Nunes, and P.~V. Moniz, ``{Inflation in a two 3-form fields scenario},'' \href{http://dx.doi.org/10.1088/1475-7516/2014/06/064}{{\em JCAP} {\bf 06} (2014)  064}, \href{http://arxiv.org/abs/1404.0211}{{\tt arXiv:1404.0211 [gr-qc]}}.

\bibitem{Barros:2015evi}
B.~J. Barros and N.~J. Nunes, ``{Three-form inflation in type II Randall-Sundrum},'' \href{http://dx.doi.org/10.1103/PhysRevD.93.043512}{{\em Phys. Rev. D} {\bf 93} (2016) no.~4, 043512}, \href{http://arxiv.org/abs/1511.07856}{{\tt arXiv:1511.07856 [astro-ph.CO]}}.

\bibitem{Wongjun:2016tva}
P.~Wongjun, ``{Perfect fluid in Lagrangian formulation due to generalized three-form field},'' \href{http://dx.doi.org/10.1103/PhysRevD.96.023516}{{\em Phys. Rev. D} {\bf 96} (2017) no.~2, 023516}, \href{http://arxiv.org/abs/1602.00682}{{\tt arXiv:1602.00682 [gr-qc]}}.

\bibitem{SravanKumar:2016biw}
K.~Sravan~Kumar, D.~J. Mulryne, N.~J. Nunes, J.~a. Marto, and P.~Vargas~Moniz, ``{Non-Gaussianity in multiple three-form field inflation},'' \href{http://dx.doi.org/10.1103/PhysRevD.94.103504}{{\em Phys. Rev. D} {\bf 94} (2016) no.~10, 103504}, \href{http://arxiv.org/abs/1606.07114}{{\tt arXiv:1606.07114 [astro-ph.CO]}}.

\bibitem{Morais:2017vlf}
J.~Morais, M.~Bouhmadi-L{\'o}pez, and J.~Marto, ``{3-Form Cosmology: Phantom Behaviour, Singularities and Interactions},'' \href{http://dx.doi.org/10.3390/universe3010021}{{\em Universe} {\bf 3} (2017) no.~1, 21}.

\bibitem{BeltranAlmeida:2018nin}
J.~P. Beltr\'an~Almeida, A.~Guarnizo, and C.~A. Valenzuela-Toledo, ``{Arbitrarily coupled $p-$forms in cosmological backgrounds},'' \href{http://dx.doi.org/10.1088/1361-6382/ab5f3c}{{\em Class. Quant. Grav.} {\bf 37} (2020) no.~3, 035001}, \href{http://arxiv.org/abs/1810.05301}{{\tt arXiv:1810.05301 [astro-ph.CO]}}.

\bibitem{Barreiro:2016aln}
T.~Barreiro, U.~Bertello, and N.~J. Nunes, ``{Screening three-form fields},'' \href{http://dx.doi.org/10.1016/j.physletb.2017.08.061}{{\em Phys. Lett. B} {\bf 773} (2017)  417--421}, \href{http://arxiv.org/abs/1610.00357}{{\tt arXiv:1610.00357 [gr-qc]}}.

\bibitem{Hell:2021wzm}
A.~Hell, ``{On the duality of massive Kalb-Ramond and Proca fields},'' \href{http://dx.doi.org/10.1088/1475-7516/2022/01/056}{{\em JCAP} {\bf 01} (2022) no.~01, 056}, \href{http://arxiv.org/abs/2109.05030}{{\tt arXiv:2109.05030 [hep-th]}}.

\bibitem{Hell:2022wci}
A.~Hell, \href{http://dx.doi.org/10.5282/edoc.30317}{{\em {The massless limit of massive gauge theories}}}.
\newblock PhD thesis, Munich U., 2022.

\bibitem{Hell:2025uoc}
A.~Hell, ``{Aspects of massive gauge fields},'' in {\em {24th Hellenic School and Workshops on Elementary Particle Physics and Gravity}}.
\newblock 5, 2025.
\newblock \href{http://arxiv.org/abs/2505.08962}{{\tt arXiv:2505.08962 [hep-th]}}.

\bibitem{Urban:2012ib}
F.~R. Urban and T.~S. Koivisto, ``{Perturbations and non-Gaussianities in three-form inflationary magnetogenesis},'' \href{http://dx.doi.org/10.1088/1475-7516/2012/09/025}{{\em JCAP} {\bf 09} (2012)  025}, \href{http://arxiv.org/abs/1207.7328}{{\tt arXiv:1207.7328 [astro-ph.CO]}}.

\bibitem{Barros:2018lca}
B.~J. Barros and F.~S.~N. Lobo, ``{Wormhole geometries supported by three-form fields},'' \href{http://dx.doi.org/10.1103/PhysRevD.98.044012}{{\em Phys. Rev. D} {\bf 98} (2018) no.~4, 044012}, \href{http://arxiv.org/abs/1806.10488}{{\tt arXiv:1806.10488 [gr-qc]}}.

\bibitem{Barros:2020ghz}
B.~J. Barros, B.~Dǎnilǎ, T.~Harko, and F.~S.~N. Lobo, ``{Black hole and naked singularity geometries supported by three-form fields},'' \href{http://dx.doi.org/10.1140/epjc/s10052-020-8178-1}{{\em Eur. Phys. J. C} {\bf 80} (2020)  617}, \href{http://arxiv.org/abs/2004.06605}{{\tt arXiv:2004.06605 [gr-qc]}}.

\bibitem{Barros:2021jbt}
B.~J. Barros, Z.~Haghani, T.~Harko, and F.~S.~N. Lobo, ``{Static spherically symmetric three-form stars},'' \href{http://dx.doi.org/10.1140/epjc/s10052-021-09105-9}{{\em Eur. Phys. J. C} {\bf 81} (2021) no.~4, 307}, \href{http://arxiv.org/abs/2101.04445}{{\tt arXiv:2101.04445 [gr-qc]}}.

\bibitem{Bouhmadi-Lopez:2020wve}
M.~Bouhmadi-L\'opez, C.-Y. Chen, X.~Y. Chew, Y.~C. Ong, and D.-H. Yeom, ``{Regular Black Hole Interior Spacetime Supported by Three-Form Field},'' \href{http://dx.doi.org/10.1140/epjc/s10052-021-09080-1}{{\em Eur. Phys. J. C} {\bf 81} (2021) no.~4, 278}, \href{http://arxiv.org/abs/2005.13260}{{\tt arXiv:2005.13260 [gr-qc]}}.

\bibitem{Bouhmadi-Lopez:2021zwt}
M.~Bouhmadi-L\'opez, C.-Y. Chen, X.~Y. Chew, Y.~C. Ong, and D.-h. Yeom, ``{Traversable wormhole in Einstein 3-form theory with self-interacting potential},'' \href{http://dx.doi.org/10.1088/1475-7516/2021/10/059}{{\em JCAP} {\bf 10} (2021)  059}, \href{http://arxiv.org/abs/2108.07302}{{\tt arXiv:2108.07302 [gr-qc]}}.

\bibitem{Normann:2017aav}
B.~D. Normann, S.~Hervik, A.~Ricciardone, and M.~Thorsrud, ``{Bianchi cosmologies with $p$-form gauge fields},'' \href{http://dx.doi.org/10.1088/1361-6382/aab3a7}{{\em Class. Quant. Grav.} {\bf 35} (2018) no.~9, 095004}, \href{http://arxiv.org/abs/1712.08752}{{\tt arXiv:1712.08752 [gr-qc]}}.

\bibitem{Bouhmadi-Lopez:2018lly}
M.~Bouhmadi-L{\'o}pez, D.~Brizuela, and I.~Garay, ``{Quantum behavior of the ''Little Sibling'' of the Big Rip induced by a three-form field},'' \href{http://dx.doi.org/10.1088/1475-7516/2018/09/031}{{\em JCAP} {\bf 09} (2018)  031}, \href{http://arxiv.org/abs/1802.05164}{{\tt arXiv:1802.05164 [gr-qc]}}.

\bibitem{Almeida:2019xzt}
J.~P.~B. Almeida, A.~Guarnizo, R.~Kase, S.~Tsujikawa, and C.~A. Valenzuela-Toledo, ``{Anisotropic inflation with coupled p{\ensuremath{-}}forms},'' \href{http://dx.doi.org/10.1088/1475-7516/2019/03/025}{{\em JCAP} {\bf 03} (2019)  025}, \href{http://arxiv.org/abs/1901.06097}{{\tt arXiv:1901.06097 [gr-qc]}}.

\bibitem{Gordin:2023nsv}
J.~E.~B. Gordin, K.~MacDevette, and J.~Bruton, ``{The dynamics of three-forms in thick branes},'' \href{http://dx.doi.org/10.1007/JHEP05(2024)061}{{\em JHEP} {\bf 05} (2024)  061}, \href{http://arxiv.org/abs/2311.14436}{{\tt arXiv:2311.14436 [hep-th]}}.

\bibitem{Dvali:2005an}
G.~Dvali, ``{Three-form gauging of axion symmetries and gravity},'' \href{http://arxiv.org/abs/hep-th/0507215}{{\tt arXiv:hep-th/0507215}}.

\bibitem{Dvali:2005zk}
G.~Dvali, ``{A Vacuum accumulation solution to the strong CP problem},'' \href{http://dx.doi.org/10.1103/PhysRevD.74.025019}{{\em Phys. Rev. D} {\bf 74} (2006)  025019}, \href{http://arxiv.org/abs/hep-th/0510053}{{\tt arXiv:hep-th/0510053}}.

\bibitem{Giudice:2019iwl}
G.~F. Giudice, A.~Kehagias, and A.~Riotto, ``{The Selfish Higgs},'' \href{http://dx.doi.org/10.1007/JHEP10(2019)199}{{\em JHEP} {\bf 10} (2019)  199}, \href{http://arxiv.org/abs/1907.05370}{{\tt arXiv:1907.05370 [hep-ph]}}.

\bibitem{Kaloper:2019xfj}
N.~Kaloper and A.~Westphal, ``{A Goldilocks Higgs},'' \href{http://dx.doi.org/10.1016/j.physletb.2020.135616}{{\em Phys. Lett. B} {\bf 808} (2020)  135616}, \href{http://arxiv.org/abs/1907.05837}{{\tt arXiv:1907.05837 [hep-th]}}.

\bibitem{Lee:2019efp}
H.~M. Lee, ``{Relaxation of Higgs mass and cosmological constant with four-form fluxes and reheating},'' \href{http://dx.doi.org/10.1007/JHEP01(2020)045}{{\em JHEP} {\bf 01} (2020)  045}, \href{http://arxiv.org/abs/1908.04252}{{\tt arXiv:1908.04252 [hep-ph]}}.

\bibitem{Lee:2019twi}
H.~M. Lee, ``{The Selfish Higgs and Reheating},'' \href{http://dx.doi.org/10.1007/JHEP04(2020)131}{{\em JHEP} {\bf 04} (2020)  131}, \href{http://arxiv.org/abs/1910.09171}{{\tt arXiv:1910.09171 [hep-ph]}}.

\bibitem{Bordin:2019fek}
L.~Bordin, F.~Cunillera, A.~Leh{\'e}bel, and A.~Padilla, ``{A natural theory of dark energy},'' \href{http://dx.doi.org/10.1103/PhysRevD.101.085012}{{\em Phys. Rev. D} {\bf 101} (2020)  085012}, \href{http://arxiv.org/abs/1912.04905}{{\tt arXiv:1912.04905 [hep-th]}}.

\bibitem{Koivisto:2009sd}
T.~S. Koivisto, D.~F. Mota, and C.~Pitrou, ``{Inflation from N-Forms and its stability},'' \href{http://dx.doi.org/10.1088/1126-6708/2009/09/092}{{\em JHEP} {\bf 09} (2009)  092}, \href{http://arxiv.org/abs/0903.4158}{{\tt arXiv:0903.4158 [astro-ph.CO]}}.

\bibitem{Germani:2009iq}
C.~Germani and A.~Kehagias, ``{P-nflation: generating cosmic Inflation with p-forms},'' \href{http://dx.doi.org/10.1088/1475-7516/2009/03/028}{{\em JCAP} {\bf 03} (2009)  028}, \href{http://arxiv.org/abs/0902.3667}{{\tt arXiv:0902.3667 [astro-ph.CO]}}.

\bibitem{Germani:2009gg}
C.~Germani and A.~Kehagias, ``{Scalar perturbations in p-nflation: the 3-form case},'' \href{http://dx.doi.org/10.1088/1475-7516/2009/11/005}{{\em JCAP} {\bf 11} (2009)  005}, \href{http://arxiv.org/abs/0908.0001}{{\tt arXiv:0908.0001 [astro-ph.CO]}}.

\bibitem{Golovnev:2009rm}
A.~Golovnev, ``{Linear perturbations in vector inflation and stability issues},'' \href{http://dx.doi.org/10.1103/PhysRevD.81.023514}{{\em Phys. Rev. D} {\bf 81} (2010)  023514}, \href{http://arxiv.org/abs/0910.0173}{{\tt arXiv:0910.0173 [astro-ph.CO]}}.

\bibitem{Golovnev:2011yc}
A.~Golovnev, ``{On cosmic inflation in vector field theories},'' \href{http://dx.doi.org/10.1088/0264-9381/28/24/245018}{{\em Class. Quant. Grav.} {\bf 28} (2011)  245018}, \href{http://arxiv.org/abs/1109.4838}{{\tt arXiv:1109.4838 [gr-qc]}}.

\bibitem{DeFelice:2025ykh}
A.~De~Felice and A.~Hell, ``{On the cosmological degrees of freedom of Proca field with non-minimal coupling to gravity},'' \href{http://dx.doi.org/10.1007/JHEP07(2025)228}{{\em JHEP} {\bf 07} (2025)  228}, \href{http://arxiv.org/abs/2503.07454}{{\tt arXiv:2503.07454 [gr-qc]}}.

\bibitem{Heisenberg:2016eld}
L.~Heisenberg, R.~Kase, and S.~Tsujikawa, ``{Beyond generalized Proca theories},'' \href{http://dx.doi.org/10.1016/j.physletb.2016.07.052}{{\em Phys. Lett. B} {\bf 760} (2016)  617--626}, \href{http://arxiv.org/abs/1605.05565}{{\tt arXiv:1605.05565 [hep-th]}}.

\bibitem{DeFelice:2016yws}
A.~De~Felice, L.~Heisenberg, R.~Kase, S.~Mukohyama, S.~Tsujikawa, and Y.-l. Zhang, ``{Cosmology in generalized Proca theories},'' \href{http://dx.doi.org/10.1088/1475-7516/2016/06/048}{{\em JCAP} {\bf 06} (2016)  048}, \href{http://arxiv.org/abs/1603.05806}{{\tt arXiv:1603.05806 [gr-qc]}}.

\bibitem{Himmetoglu:2009qi}
B.~Himmetoglu, C.~R. Contaldi, and M.~Peloso, ``{Ghost instabilities of cosmological models with vector fields nonminimally coupled to the curvature},'' \href{http://dx.doi.org/10.1103/PhysRevD.80.123530}{{\em Phys. Rev. D} {\bf 80} (2009)  123530}, \href{http://arxiv.org/abs/0909.3524}{{\tt arXiv:0909.3524 [astro-ph.CO]}}.

\bibitem{Himmetoglu:2008zp}
B.~Himmetoglu, C.~R. Contaldi, and M.~Peloso, ``{Instability of anisotropic cosmological solutions supported by vector fields},'' \href{http://dx.doi.org/10.1103/PhysRevLett.102.111301}{{\em Phys. Rev. Lett.} {\bf 102} (2009)  111301}, \href{http://arxiv.org/abs/0809.2779}{{\tt arXiv:0809.2779 [astro-ph]}}.

\bibitem{Capanelli:2024pzd}
C.~Capanelli, L.~Jenks, E.~W. Kolb, and E.~McDonough, ``{Runaway Gravitational Production of Dark Photons},'' \href{http://dx.doi.org/10.1103/PhysRevLett.133.061602}{{\em Phys. Rev. Lett.} {\bf 133} (2024) no.~6, 061602}, \href{http://arxiv.org/abs/2403.15536}{{\tt arXiv:2403.15536 [hep-th]}}.

\bibitem{Hell:2024xbv}
A.~Hell, ``{Unveiling the Inconsistency of the Proca Theory with Nonminimal Coupling to Gravity},'' \href{http://dx.doi.org/10.1093/ptep/ptae188}{{\em PTEP} {\bf 2025} (2025) no.~1, 013E01}, \href{http://arxiv.org/abs/2403.18673}{{\tt arXiv:2403.18673 [gr-qc]}}.

\bibitem{Hell:2025lgn}
A.~Hell and M.~Sasaki, ``{Accelerating Universe from Constraints},'' \href{http://arxiv.org/abs/2507.00986}{{\tt arXiv:2507.00986 [hep-th]}}.

\bibitem{Afshordi:2006ad}
N.~Afshordi, D.~J.~H. Chung, and G.~Geshnizjani, ``{Cuscuton: A Causal Field Theory with an Infinite Speed of Sound},'' \href{http://dx.doi.org/10.1103/PhysRevD.75.083513}{{\em Phys. Rev. D} {\bf 75} (2007)  083513}, \href{http://arxiv.org/abs/hep-th/0609150}{{\tt arXiv:hep-th/0609150}}.

\bibitem{Afshordi:2007yx}
N.~Afshordi, D.~J.~H. Chung, M.~Doran, and G.~Geshnizjani, ``{Cuscuton Cosmology: Dark Energy meets Modified Gravity},'' \href{http://dx.doi.org/10.1103/PhysRevD.75.123509}{{\em Phys. Rev. D} {\bf 75} (2007)  123509}, \href{http://arxiv.org/abs/astro-ph/0702002}{{\tt arXiv:astro-ph/0702002}}.

\bibitem{Brans:1961sx}
C.~Brans and R.~H. Dicke, ``{Mach's principle and a relativistic theory of gravitation},'' \href{http://dx.doi.org/10.1103/PhysRev.124.925}{{\em Phys. Rev.} {\bf 124} (1961)  925--935}.

\bibitem{Teyssandier:1983zz}
P.~Teyssandier and P.~Tourrenc, ``{The Cauchy problem for the R+R**2 theories of gravity without torsion},'' \href{http://dx.doi.org/10.1063/1.525659}{{\em J. Math. Phys.} {\bf 24} (1983)  2793}.

\bibitem{OHanlon:1972xqa}
J.~O'Hanlon, ``{Intermediate-range gravity - a generally covariant model},'' \href{http://dx.doi.org/10.1103/PhysRevLett.29.137}{{\em Phys. Rev. Lett.} {\bf 29} (1972)  137--138}.

\bibitem{Wetterich:1987fm}
C.~Wetterich, ``{Cosmology and the Fate of Dilatation Symmetry},'' \href{http://dx.doi.org/10.1016/0550-3213(88)90193-9}{{\em Nucl. Phys. B} {\bf 302} (1988)  668--696}, \href{http://arxiv.org/abs/1711.03844}{{\tt arXiv:1711.03844 [hep-th]}}.

\bibitem{Wetterich:1987fk}
C.~Wetterich, ``{Cosmologies With Variable Newton's 'Constant'},'' \href{http://dx.doi.org/10.1016/0550-3213(88)90192-7}{{\em Nucl. Phys. B} {\bf 302} (1988)  645--667}.

\bibitem{Wetterich:2013jsa}
C.~Wetterich, ``{Variable gravity Universe},'' \href{http://dx.doi.org/10.1103/PhysRevD.89.024005}{{\em Phys. Rev. D} {\bf 89} (2014) no.~2, 024005}, \href{http://arxiv.org/abs/1308.1019}{{\tt arXiv:1308.1019 [astro-ph.CO]}}.

\bibitem{Armendariz-Picon:1999hyi}
C.~Armendariz-Picon, T.~Damour, and V.~F. Mukhanov, ``{k - inflation},'' \href{http://dx.doi.org/10.1016/S0370-2693(99)00603-6}{{\em Phys. Lett. B} {\bf 458} (1999)  209--218}, \href{http://arxiv.org/abs/hep-th/9904075}{{\tt arXiv:hep-th/9904075}}.

\bibitem{Romano:2016jlz}
A.~E. Romano, ``{General background conditions for K-bounce and adiabaticity},'' \href{http://dx.doi.org/10.1140/epjc/s10052-017-4698-8}{{\em Eur. Phys. J. C} {\bf 77} (2017) no.~3, 147}, \href{http://arxiv.org/abs/1607.08533}{{\tt arXiv:1607.08533 [gr-qc]}}.

\bibitem{Lin:2017fec}
C.~Lin, J.~Quintin, and R.~H. Brandenberger, ``{Massive gravity and the suppression of anisotropies and gravitational waves in a matter-dominated contracting universe},'' \href{http://dx.doi.org/10.1088/1475-7516/2018/01/011}{{\em JCAP} {\bf 01} (2018)  011}, \href{http://arxiv.org/abs/1711.10472}{{\tt arXiv:1711.10472 [hep-th]}}.

\bibitem{Boruah:2018pvq}
S.~S. Boruah, H.~J. Kim, M.~Rouben, and G.~Geshnizjani, ``{Cuscuton bounce},'' \href{http://dx.doi.org/10.1088/1475-7516/2018/08/031}{{\em JCAP} {\bf 08} (2018)  031}, \href{http://arxiv.org/abs/1802.06818}{{\tt arXiv:1802.06818 [gr-qc]}}.

\bibitem{Quintin:2019orx}
J.~Quintin and D.~Yoshida, ``{Cuscuton gravity as a classically stable limiting curvature theory},'' \href{http://dx.doi.org/10.1088/1475-7516/2020/02/016}{{\em JCAP} {\bf 02} (2020)  016}, \href{http://arxiv.org/abs/1911.06040}{{\tt arXiv:1911.06040 [gr-qc]}}.

\bibitem{Sakakihara:2020rdy}
Y.~Sakakihara, D.~Yoshida, K.~Takahashi, and J.~Quintin, ``{Theories with limited extrinsic curvature and a nonsingular anisotropic universe},'' \href{http://dx.doi.org/10.1103/PhysRevD.102.084004}{{\em Phys. Rev. D} {\bf 102} (2020) no.~8, 084004}, \href{http://arxiv.org/abs/2005.10844}{{\tt arXiv:2005.10844 [gr-qc]}}.

\bibitem{Kim:2020iwq}
J.~L. Kim and G.~Geshnizjani, ``{Spectrum of Cuscuton Bounce},'' \href{http://dx.doi.org/10.1088/1475-7516/2021/03/104}{{\em JCAP} {\bf 03} (2021)  104}, \href{http://arxiv.org/abs/2010.06645}{{\tt arXiv:2010.06645 [gr-qc]}}.

\bibitem{Dehghani:2025udv}
A.~Dehghani, G.~Geshnizjani, and J.~Quintin, ``{Cuscuton bounce beyond the linear regime: bispectrum and strong coupling constraints},'' \href{http://dx.doi.org/10.1088/1475-7516/2025/05/026}{{\em JCAP} {\bf 05} (2025)  026}, \href{http://arxiv.org/abs/2503.01992}{{\tt arXiv:2503.01992 [hep-th]}}.

\bibitem{Moghtaderi:2025cns}
E.~Moghtaderi, B.~R. Hull, J.~Quintin, and G.~Geshnizjani, ``{How much null-energy-condition breaking can the Universe endure?},'' \href{http://dx.doi.org/10.1103/j6hp-p8hs}{{\em Phys. Rev. D} {\bf 111} (2025) no.~12, 123552}, \href{http://arxiv.org/abs/2503.19955}{{\tt arXiv:2503.19955 [gr-qc]}}.

\bibitem{Ito:2019ztb}
A.~Ito, Y.~Sakakihara, and J.~Soda, ``{Accelerating Universe with a stable extra dimension in cuscuton gravity},'' \href{http://dx.doi.org/10.1103/PhysRevD.100.063531}{{\em Phys. Rev. D} {\bf 100} (2019) no.~6, 063531}, \href{http://arxiv.org/abs/1906.10363}{{\tt arXiv:1906.10363 [gr-qc]}}.

\bibitem{Afshordi:2009tt}
N.~Afshordi, ``{Cuscuton and low energy limit of Horava-Lifshitz gravity},'' \href{http://dx.doi.org/10.1103/PhysRevD.80.081502}{{\em Phys. Rev. D} {\bf 80} (2009)  081502}, \href{http://arxiv.org/abs/0907.5201}{{\tt arXiv:0907.5201 [hep-th]}}.

\bibitem{Bhattacharyya:2016mah}
J.~Bhattacharyya, A.~Coates, M.~Colombo, A.~E. Gumrukcuoglu, and T.~P. Sotiriou, ``{Revisiting the cuscuton as a Lorentz-violating gravity theory},'' \href{http://dx.doi.org/10.1103/PhysRevD.97.064020}{{\em Phys. Rev. D} {\bf 97} (2018) no.~6, 064020}, \href{http://arxiv.org/abs/1612.01824}{{\tt arXiv:1612.01824 [hep-th]}}.

\bibitem{Gomes:2017tzd}
H.~Gomes and D.~C. Guariento, ``{Hamiltonian analysis of the cuscuton},'' \href{http://dx.doi.org/10.1103/PhysRevD.95.104049}{{\em Phys. Rev. D} {\bf 95} (2017) no.~10, 104049}, \href{http://arxiv.org/abs/1703.08226}{{\tt arXiv:1703.08226 [gr-qc]}}.

\bibitem{Cooney:2008wk}
A.~Cooney, S.~DeDeo, and D.~Psaltis, ``{Gravity with Perturbative Constraints: Dark Energy Without New Degrees of Freedom},'' \href{http://dx.doi.org/10.1103/PhysRevD.79.044033}{{\em Phys. Rev. D} {\bf 79} (2009)  044033}, \href{http://arxiv.org/abs/0811.3635}{{\tt arXiv:0811.3635 [astro-ph]}}.

\bibitem{Chagoya:2016inc}
J.~Chagoya and G.~Tasinato, ``{A geometrical approach to degenerate scalar-tensor theories},'' \href{http://dx.doi.org/10.1007/JHEP02(2017)113}{{\em JHEP} {\bf 02} (2017)  113}, \href{http://arxiv.org/abs/1610.07980}{{\tt arXiv:1610.07980 [hep-th]}}.

\bibitem{deRham:2016ged}
C.~de~Rham and H.~Motohashi, ``{Caustics for Spherical Waves},'' \href{http://dx.doi.org/10.1103/PhysRevD.95.064008}{{\em Phys. Rev. D} {\bf 95} (2017) no.~6, 064008}, \href{http://arxiv.org/abs/1611.05038}{{\tt arXiv:1611.05038 [hep-th]}}.

\bibitem{Boruah:2017tvg}
S.~S. Boruah, H.~J. Kim, and G.~Geshnizjani, ``{Theory of Cosmological Perturbations with Cuscuton},'' \href{http://dx.doi.org/10.1088/1475-7516/2017/07/022}{{\em JCAP} {\bf 07} (2017)  022}, \href{http://arxiv.org/abs/1704.01131}{{\tt arXiv:1704.01131 [hep-th]}}.

\bibitem{Lin:2017oow}
C.~Lin and S.~Mukohyama, ``{A Class of Minimally Modified Gravity Theories},'' \href{http://dx.doi.org/10.1088/1475-7516/2017/10/033}{{\em JCAP} {\bf 10} (2017)  033}, \href{http://arxiv.org/abs/1708.03757}{{\tt arXiv:1708.03757 [gr-qc]}}.

\bibitem{Iyonaga:2018vnu}
A.~Iyonaga, K.~Takahashi, and T.~Kobayashi, ``{Extended Cuscuton: Formulation},'' \href{http://dx.doi.org/10.1088/1475-7516/2018/12/002}{{\em JCAP} {\bf 12} (2018)  002}, \href{http://arxiv.org/abs/1809.10935}{{\tt arXiv:1809.10935 [gr-qc]}}.

\bibitem{Pajer:2018egx}
E.~Pajer and D.~Stefanyszyn, ``{Symmetric Superfluids},'' \href{http://dx.doi.org/10.1007/JHEP06(2019)008}{{\em JHEP} {\bf 06} (2019)  008}, \href{http://arxiv.org/abs/1812.05133}{{\tt arXiv:1812.05133 [hep-th]}}.

\bibitem{Gao:2019twq}
X.~Gao and Z.-B. Yao, ``{Spatially covariant gravity theories with two tensorial degrees of freedom: the formalism},'' \href{http://dx.doi.org/10.1103/PhysRevD.101.064018}{{\em Phys. Rev. D} {\bf 101} (2020) no.~6, 064018}, \href{http://arxiv.org/abs/1910.13995}{{\tt arXiv:1910.13995 [gr-qc]}}.

\bibitem{Grall:2019qof}
T.~Grall, S.~Jazayeri, and E.~Pajer, ``{Symmetric Scalars},'' \href{http://dx.doi.org/10.1088/1475-7516/2020/05/031}{{\em JCAP} {\bf 05} (2020)  031}, \href{http://arxiv.org/abs/1909.04622}{{\tt arXiv:1909.04622 [hep-th]}}.

\bibitem{Mukohyama:2019unx}
S.~Mukohyama and K.~Noui, ``{Minimally Modified Gravity: a Hamiltonian Construction},'' \href{http://dx.doi.org/10.1088/1475-7516/2019/07/049}{{\em JCAP} {\bf 07} (2019)  049}, \href{http://arxiv.org/abs/1905.02000}{{\tt arXiv:1905.02000 [gr-qc]}}.

\bibitem{DeFelice:2020eju}
A.~De~Felice, A.~Doll, and S.~Mukohyama, ``{A theory of type-II minimally modified gravity},'' \href{http://dx.doi.org/10.1088/1475-7516/2020/09/034}{{\em JCAP} {\bf 09} (2020)  034}, \href{http://arxiv.org/abs/2004.12549}{{\tt arXiv:2004.12549 [gr-qc]}}.

\bibitem{Iyonaga:2020bmm}
A.~Iyonaga, K.~Takahashi, and T.~Kobayashi, ``{Extended Cuscuton as Dark Energy},'' \href{http://dx.doi.org/10.1088/1475-7516/2020/07/004}{{\em JCAP} {\bf 07} (2020)  004}, \href{http://arxiv.org/abs/2003.01934}{{\tt arXiv:2003.01934 [gr-qc]}}.

\bibitem{Pookkillath:2021gdp}
M.~C. Pookkillath, ``{Minimally Modified Gravity Fitting Planck Data Better Than {\ensuremath{\Lambda}}CDM},'' \href{http://dx.doi.org/10.1134/S1063772921100279}{{\em Astron. Rep.} {\bf 65} (2021) no.~10, 1021--1025}.

\bibitem{DeFelice:2022uxv}
A.~De~Felice, K.-i. Maeda, S.~Mukohyama, and M.~C. Pookkillath, ``{Comparison of two theories of Type-IIa minimally modified gravity},'' \href{http://dx.doi.org/10.1103/PhysRevD.106.024028}{{\em Phys. Rev. D} {\bf 106} (2022) no.~2, 024028}, \href{http://arxiv.org/abs/2204.08294}{{\tt arXiv:2204.08294 [gr-qc]}}.

\bibitem{Ganz:2022iiv}
A.~Ganz, ``{Dynamical dark energy in minimally modified gravity},'' \href{http://dx.doi.org/10.1088/1475-7516/2022/08/074}{{\em JCAP} {\bf 08} (2022)  074}, \href{http://arxiv.org/abs/2203.12358}{{\tt arXiv:2203.12358 [gr-qc]}}.

\bibitem{Mylova:2023ddj}
M.~Mylova and N.~Afshordi, ``{Effective cuscuton theory},'' \href{http://dx.doi.org/10.1007/JHEP04(2024)144}{{\em JHEP} {\bf 04} (2024)  144}, \href{http://arxiv.org/abs/2312.06066}{{\tt arXiv:2312.06066 [hep-th]}}.

\bibitem{Bazeia:2025mzr}
D.~Bazeia, J.~D. Dantas, and S.~S. da~Costa, ``{Cuscuton-like contribution to dark energy evolution},'' \href{http://dx.doi.org/10.1140/epjc/s10052-025-13855-1}{{\em Eur. Phys. J. C} {\bf 85} (2025)  196}, \href{http://arxiv.org/abs/2501.14909}{{\tt arXiv:2501.14909 [astro-ph.CO]}}.

\bibitem{Mukhanov:2005sc}
V.~Mukhanov, \href{http://dx.doi.org/10.1017/CBO9780511790553}{{\em {Physical Foundations of Cosmology}}}.
\newblock Cambridge University Press, Oxford, 2005.

\bibitem{Bardeen:1980kt}
J.~M. Bardeen, ``{Gauge Invariant Cosmological Perturbations},'' \href{http://dx.doi.org/10.1103/PhysRevD.22.1882}{{\em Phys. Rev. D} {\bf 22} (1980)  1882--1905}.

\bibitem{Dicke:1961gz}
R.~H. Dicke, ``{Mach's principle and invariance under transformation of units},'' \href{http://dx.doi.org/10.1103/PhysRev.125.2163}{{\em Phys. Rev.} {\bf 125} (1962)  2163--2167}.

\bibitem{Vainshtein:1972sx}
A.~I. Vainshtein, ``{To the problem of nonvanishing gravitation mass},'' \href{http://dx.doi.org/10.1016/0370-2693(72)90147-5}{{\em Phys. Lett. B} {\bf 39} (1972)  393--394}.

\bibitem{Deffayet:2001uk}
C.~Deffayet, G.~R. Dvali, G.~Gabadadze, and A.~I. Vainshtein, ``{Nonperturbative continuity in graviton mass versus perturbative discontinuity},'' \href{http://dx.doi.org/10.1103/PhysRevD.65.044026}{{\em Phys. Rev. D} {\bf 65} (2002)  044026}, \href{http://arxiv.org/abs/hep-th/0106001}{{\tt arXiv:hep-th/0106001}}.

\bibitem{Gruzinov:2001hp}
A.~Gruzinov, ``{On the graviton mass},'' \href{http://dx.doi.org/10.1016/j.newast.2004.12.001}{{\em New Astron.} {\bf 10} (2005)  311--314}, \href{http://arxiv.org/abs/astro-ph/0112246}{{\tt arXiv:astro-ph/0112246}}.

\bibitem{Dvali:2006su}
G.~Dvali, ``{Predictive Power of Strong Coupling in Theories with Large Distance Modified Gravity},'' \href{http://dx.doi.org/10.1088/1367-2630/8/12/326}{{\em New J. Phys.} {\bf 8} (2006)  326}, \href{http://arxiv.org/abs/hep-th/0610013}{{\tt arXiv:hep-th/0610013}}.

\bibitem{Dvali:2007kt}
G.~Dvali, S.~Hofmann, and J.~Khoury, ``{Degravitation of the cosmological constant and graviton width},'' \href{http://dx.doi.org/10.1103/PhysRevD.76.084006}{{\em Phys. Rev. D} {\bf 76} (2007)  084006}, \href{http://arxiv.org/abs/hep-th/0703027}{{\tt arXiv:hep-th/0703027}}.

\bibitem{Hell:2021oea}
A.~Hell, ``{The strong couplings of massive Yang-Mills theory},'' \href{http://dx.doi.org/10.1007/JHEP03(2022)167}{{\em JHEP} {\bf 03} (2022)  167}, \href{http://arxiv.org/abs/2111.00017}{{\tt arXiv:2111.00017 [hep-th]}}.

\bibitem{Hell:2023mph}
A.~Hell, D.~Lust, and G.~Zoupanos, ``{On the degrees of freedom of R$^{2}$ gravity in flat spacetime},'' \href{http://dx.doi.org/10.1007/JHEP02(2024)039}{{\em JHEP} {\bf 02} (2024)  039}, \href{http://arxiv.org/abs/2311.08216}{{\tt arXiv:2311.08216 [hep-th]}}.

\bibitem{Hell:2025pso}
A.~Hell, \href{http://dx.doi.org/10.1007/978-3-032-01090-2}{{\em {The Massless Limit of Massive Gauge Theories}: {To the Strong Coupling and Beyond}}}.
\newblock Springer Theses. Springer Cham, 7, 2025.

\bibitem{Quevedo:1996uu}
F.~Quevedo and C.~A. Trugenberger, ``{Phases of antisymmetric tensor field theories},'' \href{http://dx.doi.org/10.1016/S0550-3213(97)00337-4}{{\em Nucl. Phys. B} {\bf 501} (1997)  143--172}, \href{http://arxiv.org/abs/hep-th/9604196}{{\tt arXiv:hep-th/9604196}}.

\end{thebibliography}\endgroup

\end{document}